\documentclass{article}
\usepackage[margin=1in]{geometry}
\usepackage{amsmath, amssymb}
\usepackage{placeins}
\usepackage{graphicx}
\usepackage{booktabs}
\usepackage{subcaption}
\usepackage{multirow}
\usepackage{authblk}
\usepackage{natbib}
\usepackage{hyperref}
\hypersetup{colorlinks=true, linkcolor=blue, citecolor=blue, urlcolor=blue}

\makeatletter

\newcommand{\authbreaksmall}{\protect\\[0ex]}
\newcommand{\authbreak}{\protect\\[2ex]}
\makeatother

\title{Explaining Surface Layer Theory Departures in Marine Flux Profiles with Data-Driven Discovery}

\author[1]{Jack Foxabbott\thanks{Corresponding author: jack@leap-labs.com}}
\author[1]{Leo Mckee-Reid}
\author[1]{Andrew Cusick}
\author[1]{Robbie McCorkell}
\author[1]{Jugal Patel}
\author[1]{\authbreaksmall Jamie Rumbelow}
\author[1]{Jessica Rumbelow}
\author[1]{Zohreh Shams}
\author[1]{Arush Tagade}
\author[2]{\authbreak Patrick Hawbecker}
\author[2]{Sue Ellen Haupt}

\affil[1]{Leap Laboratories, London, United Kingdom}
\affil[2]{NSF National Center for Atmospheric Research (NSF NCAR), Boulder, Colorado, United States}

\date{}

\begin{document}

\maketitle

\begin{abstract}
    Monin--Obukhov Similarity Theory (MOST), which underpins nearly all bulk estimates of surface fluxes in the atmospheric surface layer, assumes monotonic wind profiles and vertically uniform momentum and heat fluxes. Here, we show that conditions frequently arise in coastal marine settings where these assumptions do not hold. Using flux measurements from the Coastal Land-Air-Sea Interaction (CLASI) project's Air-Sea Interaction Spar (ASIS) buoys with wind and flux measurements at typically ~3m and ~5m above the sea surface, we find that wind speed decreases with height in nearly 20\% of observations, and that large vertical gradients in sensible heat flux occur near the surface, contrary to what would be predicted by MOST. Both anomalies are strongly modulated by coastal proximity and wind direction, with the highest occurrence rates near shore under offshore winds. These patterns were found using the Discovery Engine, a general-purpose automated system for scientific discovery, which identifies complex relationships in data without prior hypotheses. The Discovery Engine uncovered three distinct mechanisms responsible for breakdowns in MOST assumptions: internal boundary layers formed by offshore continental flow, wave-driven wind jets associated with high wave age, and thermally stable boundary layers over cold sea surfaces where warm, moist air overlies cooler water far from shore. These findings highlight the limitations of current flux algorithms and suggest directions for improved parameterisation in coastal and open-ocean conditions.
\end{abstract}

\section{Introduction}

Air–sea fluxes of momentum, heat, and moisture play a central role in atmosphere–ocean dynamics. They govern boundary-layer structure, contribute to weather and climate variability, and must be parameterised in numerical models at all scales. Despite their importance, direct flux measurements remain sparse, and most estimates rely on simplified models of surface-layer behaviour.

The prevailing framework for such models is Monin--Obukhov Similarity Theory (MOST), which describes the vertical structure of the atmospheric surface layer under assumptions of stationarity, horizontal homogeneity, and constant flux with height \citep{monin1954basic, businger1971flux}. MOST is foundational for parameterizing surface fluxes in atmospheric and climate models, including the COARE family \citep{fairall2003bulk}, the Global Forecast System (GFS) \citep{ncep2003gfs}, the Weather Research and Forecasting Model (WRF) \citep{wrf2008, wrf2019}, climate models including CLM \citep{dai2003common} and JULES \citep{julespart1, julespart2}, and the Coupled Model Intercomparison Project (CMIP) models \citep{cmip1, cmip2, cmip3}.
However, MOST was not developed for offshore surface layers, and its assumptions often break down in complex marine environments shaped by coastal transitions, wave dynamics, or stratification.

Recent field campaigns have revealed consistent and systematic deviations from MOST predictions. In particular, instances in which wind speed decreases with height at low levels (under 10 m)—here termed negative wind speed gradients—have been reported in both observational and modelling studies \citep{WaveDrivenWindJetsintheMarineAtmosphericBoundaryLayer, InfluenceofSwellonMarineSurfaceLayerStructure}. Similarly, large vertical gradients in sensible heat flux have been observed under strongly stable conditions \citep{mahrt1998stratified, mahrt2014stably}, challenging the assumption of vertically uniform fluxes. Throughout this study, these gradients are evaluated between measurement heights of approximately 3m and 5m above the sea surface. These anomalies are not easily explained by measurement noise or surface-layer turbulence alone and suggest the presence of distinct dynamical regimes outside the scope of MOST.

To systematically investigate these departures from MOST assumptions, we analyse flux measurements from the Coastal Land-Air-Sea Interaction (CLASI) project's Air-Sea Interaction Spar (ASIS) buoys \citep{CLASICoordinatingInnovativeObservationsandModelingtoImproveCoastalEnvironmentalPredictionSystems} positioned within 5 metres of the sea surface, with flux and wind measurements typically at $\sim$3~m and $\sim$5~m above sea level, spanning multiple coastal and open-ocean deployments. Our focus is on two instances in which observations fall outside the assumptions underpinning MOST: (1) negative wind speed gradients, and (2) large vertical gradients in sensible heat flux. We treat these as empirical markers of surface-layer disruption and ask: when and where do they occur, and what environmental conditions predict their presence?

This work is the first application of the Discovery Engine \citep{leap2025discovery} to meteorological data. Rather than performing laborious manual data analysis to investigate pre-specified hypotheses, which potentially encode biases and assumptions that limit possible insights, this system enables a data-first approach by automatically extracting interpretable patterns in the data linked to outcomes of interest -- in this case, identifying environmental regimes that strongly predict surface layer anomalies. The resulting findings are both statistically robust and physically interpretable, enabling researchers to uncover behaviour that lies outside the scope of classical surface layer theory, grounded in empirical structure. This approach allows us to move beyond association and towards data-driven hypotheses about the mechanisms responsible for breakdowns in the MOST assumptions.

Using this system, we found three distinct regimes predictive of MOST departures: (i) coastal internal boundary layers formed during offshore flow, (ii) wave-driven wind jets associated with high wave age, and (iii) thermally stable boundary layers over cold sea surfaces far from shore. Each regime reflects a physically plausible mechanism that alters turbulence, modifies flux profiles, or suppresses mixing, leading to systematic breakdowns of MOST.

This paper proceeds as follows. Section~\ref{sec:background} reviews the assumptions of surface-layer theory and previous literature on its limitations. Section~\ref{sec:data} describes the dataset and feature construction. Section~\ref{sec:disco} outlines the method implemented by the Discovery Engine. Section~\ref{sec:results} presents our main empirical results, and Section~\ref{discussion} interprets these findings in light of known mechanisms. We conclude in Section~\ref{sec:conclusion} with implications for flux modelling and future observational strategies.

\section{Background}\label{sec:background}

\subsection{Surface Layer Theory}

Monin--Obukhov Similarity Theory (MOST) is the dominant framework used to describe the structure of the atmospheric surface layer, the lowest tens of meters of the atmosphere where turbulent fluxes of momentum, heat, and moisture are assumed to be vertically constant. Developed in the 1950s, MOST provides scaling relationships for how mean profiles of wind speed, temperature, and humidity vary with height, depending on surface fluxes and atmospheric stability \citep{monin1954basic, businger1971flux}.

MOST assumes that turbulence is the primary transport mechanism in the surface layer, and that the flow is horizontally homogeneous and stationary. Under these assumptions, the vertical profile of wind speed $u(z)$ is described by the equation:

\begin{equation}
    u(z) = \frac{u_*}{\kappa} \left[ \ln\left( \frac{z}{z_0} \right) - \psi_m\left( \frac{z}{L} \right) \right],
\end{equation}

where $u(z)$ is the mean horizontal wind speed at height $z$ above the surface; $u_*$ is the friction velocity, a measure of surface shear stress; $\kappa \approx 0.4$ is the von Kármán constant; $z_0$ is the roughness length of the surface (related to surface drag); $L$ is the Monin--Obukhov length, a stability parameter; and $\psi_m(z/L)$ is a stability correction function for momentum.

In neutral conditions ($L \to \infty$), the stability correction $\psi_m$ vanishes, and the wind profile simplifies to a logarithmic form:

\begin{equation}
    u(z) = \frac{u_*}{\kappa} \ln\left( \frac{z}{z_0} \right).
\end{equation}

The Monin--Obukhov length $L$ defines the relative importance of thermal and mechanical forcing in the surface layer and is given by:

\begin{equation}
    L = -\frac{u_*^3}{\kappa (g/T_0) \overline{w'\theta'}},
\end{equation}

where $g$ is the gravitational acceleration, $T_0$ is a reference temperature (often the surface air temperature), and $\overline{w'\theta'}$ is the kinematic heat flux. When $L < 0$, conditions are unstable (buoyant turbulence dominates); when $L > 0$, the atmosphere is stably stratified (buoyancy suppresses turbulence).


MOST predicts that wind speed increases monotonically with height, that fluxes of momentum and heat are roughly constant with height within the surface layer, and that stability-corrected logarithmic profiles can be used to extrapolate wind or temperature between measurement levels. These properties form the theoretical basis for many surface flux estimation techniques, including bulk aerodynamic algorithms.

Although MOST has been widely applied and validated over idealised land \citep{businger1971flux, kaimal1976turbulence, lee2020evaluation} and ocean \citep{Venora2009, li2012monin, holtslag2015validation} surfaces, its assumptions may not hold in more complex marine environments -- especially those influenced by wave dynamics, coastal transitions, or stable stratification. In such settings, deviations from the expected log-linear structure have been noted in previous studies \citep{WaveDrivenWindJetsintheMarineAtmosphericBoundaryLayer, InfluenceofSwellonMarineSurfaceLayerStructure, mahrt1998stratified, mahrt2014stably}, motivating continued investigation into the limits of MOST and the potential need for more adaptive or data-driven approaches.

\subsection{Bulk Flux Algorithms}

Bulk flux algorithms are the primary tools used to estimate air–sea exchanges of momentum, heat, and moisture in numerical models \citep{JIANG2024107486, reeves2021ocean}, satellite products \citep{TANG2024104662}, and field campaigns \citep{cronin2019air}. These algorithms infer turbulent fluxes from time-averaged meteorological measurements, such as wind speed, temperature, humidity, and sea surface temperature, by applying similarity-based scaling laws to relate these variables to surface stress and heat exchange.

Nearly all modern bulk flux algorithms are founded on MOST, which, in the surface layer (typically defined as the lowest 10\% of the atmospheric boundary layer), assumes horizontally homogeneous, stationary flow with vertically constant fluxes and a balance between shear production and turbulent dissipation. Under these assumptions, wind and scalar profiles are expected to follow stability-corrected logarithmic forms, allowing the fluxes to be inferred from observed profiles.

Prominent examples include the COARE algorithm family \citep{fairall2003bulk}, widely used for open-ocean flux estimation; flux schemes within operational models such as the Global Forecast System (GFS) \citep{ncep2003gfs} and the Weather Research and Forecasting (WRF) model \citep{wrf2008, wrf2019}; and parameterisations embedded in Earth system models such as CLM \citep{dai2003common} and JULES \citep{julespart1, julespart2}, as well as the suite of models participating in CMIP \citep{cmip1, cmip2, cmip3}. These algorithms differ in their treatment of specific processes—for example, the inclusion of sea-state-dependent roughness, gustiness corrections, or high-wind adjustments—but all share the fundamental reliance on MOST.

While these flux algorithms are used in a wide range of conditions, particularly over open oceans and uniform land surfaces, their accuracy in coastal zones, wave-dominated regions, or strongly stratified boundary layers is less well understood. Importantly, all such algorithms retain the assumption of vertically uniform fluxes and monotonic wind profiles within the surface layer. This means they may systematically misrepresent fluxes in conditions where these assumptions break down, such as in the presence of internal boundary layers \citep{stableinternalboundary, scalingroughness}, wave-driven wind jets \citep{WaveDrivenWindJetsintheMarineAtmosphericBoundaryLayer, InfluenceofSwellonMarineSurfaceLayerStructure}, or stable stratification \citep{mahrt1998stratified, mahrt2014stably}.

Our study investigates such conditions, highlighting the need for either refinement of existing bulk flux algorithms or the development of alternative parameterisations that better capture the diversity of surface-layer structures encountered in real-world marine environments.

\subsection{Observations Incompatible with MOST} \label{sec:violations-of-interest}
Although MOST allows for stability-dependent curvature in vertical profiles, certain phenomena are unlikely under its assumptions, owing to the fact that it was not originally intended to model offshore surface layers. Two such phenomena are of particular interest in this study.

First, MOST predicts that wind speed should increase monotonically with height within the surface layer. This is a direct consequence of the assumed balance between shear production and turbulent dissipation in a steady, horizontally homogeneous surface layer. Instances in which wind speed decreases with height, referred to here as negative wind speed gradients, are not described by the log-profile formulation and suggest the influence of additional mechanisms beyond those captured by the theory.

Second, MOST assumes that surface fluxes of momentum and heat are roughly constant with height within the surface layer. In practice, some variability is expected due to local turbulence and measurement uncertainty, but large vertical gradients in fluxes, particularly in the sensible heat flux, indicate conditions where the constant-flux assumption no longer applies. 

These two departures from MOST assumptions—negative wind speed gradients and large (defined as greater than 0.01 $K \cdot s^{-1}$) sensible heat flux gradients—form the core anomalies investigated in this study. We stress that these anomalies refer specifically to gradients measured between the low-level heights of about 3 m and 5 m. In line with MOST’s assumption of vertically uniform fluxes, we expect sensible heat flux to vary by no more than $\sim$10\% with height. However, because sensible heat flux can take both positive and negative values, percentage changes are often ill-defined. While some studies separate positive and negative gradients, we focus on the absolute gradient to provide a concise summary of all substantial deviations from the constant-flux assumption. This choice reflects the similarity of our findings for positive and negative gradients individually, as well as our aim to present the results clearly without unnecessary duplication. We treat them both as empirical indicators of surface layer dynamics that fall outside the scope of MOST and related similarity-based frameworks.

\subsection{Known Exceptions to Log-Profile Assumptions}
While Monin--Obukhov Similarity Theory (MOST) has been widely applied in marine boundary layer research, numerous studies have documented conditions under which its assumptions break down. These include specific mechanisms that introduce deviations from the expected log-linear wind profiles, particularly in coastal and wave-affected environments. In this section, we summarise five well-known physical scenarios that can lead to departures from the standard surface-layer structure.

\subsubsection{Wave-Driven Wind Jets} \label{sec:wave-driven-wind-jets}
It is well documented that swell-generated waves often impart momentum upward into the atmosphere, particularly when the waves travel faster than the local wind \citep{WaveDrivenWindJetsintheMarineAtmosphericBoundaryLayer, WaveInducedWindintheMarineBoundaryLayer, EvidenceofEnergyandMomentumFluxfromSwelltoWind}. Large-eddy simulations consistently show a low-level jet forming near the surface when swell dominates, with wind speed peaking at heights as low as 2.93m (though more frequently around 30m) before decreasing with altitude \citep{ImpactofSwellonAirSeaMomentumFluxandMarineBoundaryLayerunderLowWindConditions, InfluenceofSwellonMarineSurfaceLayerStructure}. These swell-driven wind jets represent a form of upward momentum flux that is not accounted for by traditional MOST-based turbulence scaling.

\subsubsection{Flow Separation and Sheltering}
In the presence of steep or breaking waves, the airflow near the surface can experience separation, forming recirculation zones or eddies in the lee of wave crests \citep{LargeEddySimulationsandObservationsofAtmosphericMarineBoundaryLayersaboveNonequilibriumSurfaceWaves, reuletal}. These sheltered zones are associated with locally reduced wind speeds and altered shear stress, particularly within the lowest few metres above the surface \citep{Yang_Shen_2017, StructureoftheAirflowaboveSurfaceWaves}. Flow separation disrupts the vertical continuity of momentum transfer and leads to profile shapes that deviate significantly from the idealised logarithmic form \citep{WaveBoundaryLayerTurbulenceoverSurfaceWavesinaStronglyForcedCondition, TheCoupledBoundaryLayersandAirSeaTransferExperimentinLowWinds}.

\subsubsection{Sea Spray and Spray-Layer Wind Maxima}
Under high-wind conditions, the production of sea spray introduces an additional pathway for momentum and heat exchange between the ocean and atmosphere. Spray droplets can carry momentum upward \citep{ParameterizationsofSeaSprayImpactontheAirSeaMomentumandHeatFluxes}, contributing to vertical fluxes in a manner not captured by MOST. As the droplets are accelerated by the wind and fall back toward the surface, they transfer momentum to the ocean and influence the total air–sea momentum flux. These processes are concentrated within a shallow layer near the surface where spray is abundant, and they can modify the vertical wind structure, although direct evidence for wind speed maxima within this spray layer is limited in the existing literature \citep{10.1063/PT.3.3363, EffectsofSeaSprayonLargeScaleClimaticFeaturesovertheSouthernOcean, typhoonmolave}.

\subsubsection{Thermal Stratification and Stable Layers}\label{sec:thermal-stratification}

Strongly stable stratification suppresses turbulent mixing, particularly during nighttime cooling or in environments where warm air flows over relatively colder water \citep{stull2012introduction, garratt1994atmospheric}. In such cases, the downward transfer of momentum is inhibited, and the near-surface flow can become weakly coupled to the layers above \citep{mahrt1998stratified, mahrt1999stratified}. Within a stable boundary layer, characterised by negative sensible heat flux, wind speed typically increases with height up to a low-level jet; however, additional processes, such as the presence of internal boundary layers, coastal transitions, or wave effects, can further disrupt the profile. These can lead to situations where wind speed decreases with height or where vertical fluxes vary sharply over small vertical distances, particularly near transitions between distinct air masses or above the stable layer itself \citep{mahrt2014stably}.


\subsubsection{Coastal Internal Boundary Layers} \label{sec:internal-boundary-layers}
When air flows from land to sea, the abrupt change in surface properties—such as roughness length, thermal inertia, and moisture availability—drives the formation of an internal boundary layer (IBL) as the flow adjusts to marine conditions \citep{stableinternalboundary, scalingroughness}. During this transition, vertical profiles of wind, temperature, and humidity often develop complex structures, including non-monotonic wind speed gradients and vertically varying fluxes \citep{scalingofthestable, mesoscale}. These IBLs are particularly prevalent in coastal regions and frequently fall outside the assumptions of MOST, challenging the applicability of bulk flux formulations. 
Data from the CLASI campaign are being used to investigate these processes \citep{CLASICoordinatingInnovativeObservationsandModelingtoImproveCoastalEnvironmentalPredictionSystems}, and our work provides evidence that internal boundary layers are an important mechanism of surface-layer theory breakdown in coastal zones, particularly under offshore flow regimes.


\section{Data}\label{sec:data}
\subsection{CLASI-ASIS Dataset}
The COASTAL Land-Air-Sea Interaction (CLASI) project is an Office of Naval Research Departmental Research Initiative that aims to collect valuable low-level measurements of the offshore environment~\citep{CLASICoordinatingInnovativeObservationsandModelingtoImproveCoastalEnvironmentalPredictionSystems}.
Four deployments were conducted within the CLASI project including three in the Monterey Bay, CA, USA (June-July, 2021; August-September, 2021; and August, 2022) and one in Destin, FL, USA (January-February, 2023).
Within each deployment, eight Air–Sea Interaction Spar (ASIS) buoys were deployed that measured heat and momentum fluxes between 3 and 5 meters above the ocean surface.
Several of these ASIS buoys collected measurements at two heights (typically at close to 3~m and 5~m).
These buoys comprise the data that are used in this study.
Along with flux data, the buoys also collect air and sea temperature, relative humidity, wind speed and direction, wave height and phase speed, and pressure.
The data are averaged over 10~minute intervals.
One of the main goals of the CLASI project and field campaigns is to use the collected data to develop a new surface layer model for the offshore environment citing issues with the currently used MOST~\citep{CLASICoordinatingInnovativeObservationsandModelingtoImproveCoastalEnvironmentalPredictionSystems}.

\subsection{Feature Engineering}

We constructed a set of engineered features to summarise the vertical structure, surface forcing, and environmental context of each observation. These features fall into three main categories: profile-derived gradients and means, wave-based descriptors, and categorical encodings of external conditions.

\subsubsection{Vertical Gradients and Mean Profiles}

For all variables with multi-level vertical measurements, we computed two key summary statistics:

\begin{itemize}
    \item \textbf{Vertical Gradient:} The difference between the highest and lowest values divided by the height difference. This was applied to wind speed, sensible heat flux, and friction velocity.
    \item \textbf{Vertical Mean:} The arithmetic mean across all available measurement levels. This was computed for wind speed, temperature, specific humidity, sensible heat flux, and friction velocity.
\end{itemize}

For wind direction, we calculated a circular mean to correctly average angles over the vertical profile.

\subsubsection{Wave-Derived Features}

We included two wave-based descriptors that have been linked to air–sea interaction regimes:

\begin{itemize}
    \item \textbf{Wave Age:} Defined as the ratio of wave phase speed to wind speed. Lower wave age values typically indicate younger, wind-driven seas. Higher wave age is typical of swell.
    \item \textbf{Wave Steepness:} Calculated as
    \[
    \text{Steepness} = \frac{g H}{2\pi c^2},
    \]
    where \( H \) is the wave height, \( c \) is the wave phase speed, and \( g \) is the gravitational constant. Higher steepness values reflect greater nonlinearity and potential for wave breaking.
\end{itemize}

\subsubsection{Thermal Gradient}

To represent thermal forcing across the air–sea interface, we introduced:

\begin{itemize}
    \item \textbf{T\_gradient:} The difference between the near-surface air temperature and sea surface temperature.
\end{itemize}

\subsubsection{Categorical and Geometric Features}

To capture the broader environmental setting, we included:

\begin{itemize}
    \item \textbf{Wind-from-land:} A binary indicator for whether the wind direction implied offshore flow (i.e., from land to sea).
    \item \textbf{Time of Day:} Categorised into four bins—\textit{morning}, \textit{afternoon}, \textit{evening}, and \textit{night}—based on local solar time.
    \item \textbf{Distance to Land:} The shortest distance from the buoy to the nearest coastline, computed using geodesic buffering over a high-resolution landmask.
    \item \textbf{Wind Speed Gradient:} A binary indicator for whether the wind speed gradient is negative.
\end{itemize}

These features were used as inputs to the Discovery Engine \citep{leap2025discovery} and statistical analyses throughout the study.

\section{The Discovery Engine}\label{sec:disco}

The Discovery Engine is an end-to-end system for scientific discovery, combining machine learning with state-of-the-art interpretability to uncover complex patterns in structured datasets. Its aim is to shift the scientific process from hypothesis-driven exploration to a data-first paradigm, enabling faster, broader, and less biased discovery across domains~\citep{leap2025discovery}. Given the complex, nonlinear relationships underlying the CLASI-ASIS data, the analysis we wished to perfect necessitated the use of a tool like the Discovery Engine. Not knowing a priori the best model for the job required an AutoML model search, and the Discovery Engine's AutoML component has been shown to output state-of-the-art models in tasks spanning a number of scientific domains \citep{foxabbott2025benchmarkingdiscoveryengine}. Furthermore, traditional methods for machine learning interpretability, such as LIME \citep{ribeiro2016whyitrustyou} and SHAP \citep{lundberg2017unifiedapproachinterpretingmodel} yield local feature-importance estimates, but are not easily translated into meaningful patterns. The Discovery Engine extracts such patterns autonomously, making it perfect for the task at hand: understanding the underlying mechanisms behind departures from MOST assumptions in the CLASI-ASIS data. It implements the following steps in sequence, automatically:

\subsection{Automated Preprocessing}

First, automatic data ingestion and preprocessing was performed. Preprocessing steps were selected heuristically based on the data characteristics, and required no manual intervention beyond some initial feature engineering, outlined in Section~\ref{sec:data}. In this case, continuous features were scaled to have mean 0 and standard deviation 1, the column mean was imputed into missing values in continuous columns, and categorical columns were one-hot encoded. This design drastically reduced the time and expertise required to prepare this kind of scientific data for analysis.

\subsection{AutoML Model Search}

Following preprocessing, the Discovery Engine automatically trained a range of machine learning models on the target variable of interest: in this case, the binary wind speed gradient indicator, and absolute sensible heat flux gradient. These included both ensemble methods and neural architectures, all selected and tuned automatically. Each model underwent hyperparameter search, early stopping, and overfitting detection, with evaluation on holdout validation and test sets to ensure robustness. The best-performing model was selected based on generalisation performance, ensuring that extracted patterns reflect real structure in the data rather than noise. For this data, the system trained 103 models for each target, including linear or logistic regression, random forests, support vector machines, XGBoost, and neural networks. For both targets, the best-performing models was XGBoost, with \texttt{max\_depth} parameter 6 for absolute sensible heat flux gradient and 9 for binary wind speed gradient. The model performance statistics can be found in Table~\ref{tab:model-performance}.

\begin{table}[ht]
\centering
\caption{Train and test performance of the regression and classification models. The regression model predicts the absolute sensible heat flux gradient using $R^2$, while the classification model predicts the binary wind speed gradient using AUC.}\label{tab:model-performance}
\begin{tabular}{l l l c c}
\toprule
\textbf{Model Type} & \textbf{Target} & \textbf{Metric} & \textbf{Train Score} & \textbf{Test Score} \\
\midrule
Regression & Absolute sensible heat flux gradient & MAE & 0.000824 & 0.001392 \\
Classification & Binary wind speed gradient & AUC & 1 & 0.798001 \\
\bottomrule
\end{tabular}
\end{table}

\subsection{Interpretability and Pattern Extraction}

Once trained, these models were analysed using interpretability methods, to systematically extract conditional rules and non-linear relationships between features and outcomes. These patterns were validated against the original dataset to distinguish empirical findings from model extrapolations. The system categorises each insight as either a discovery (present in the data and statistically supported) or a hypothesis (an extrapolated prediction requiring further verification), enabling rigorous assessment of novel patterns. In our case, around 20 discoveries were made for each of the targets.

\subsection{Outputs and Reporting}

The output of the system is these patterns, ranked by statistical strength and novelty, and contextualised with respect to existing scientific literature. The final output includes: a LaTeX report rendered as a PDF, complete with citations; reproducible code artefacts; and the best predictive model -- allowing findings to be shared, validated, and extended in future work.

In this paper we present the most interesting discoveries made by Discovery Engine, providing evidence that coastal internal boundary layers (Section~\ref{sec:internal-boundary-layers}) cause breakdowns of the MOST assumptions near the coast, and that wave-driven wind jets (Section~\ref{sec:wave-driven-wind-jets}) and thermal stratification (Section~\ref{sec:thermal-stratification}) contribute to these breakdowns further out at sea.

\section{Results}\label{sec:results}

\subsection{Primary Discovery: Offshore Winds Predict Negative Wind Speed Gradients}

\begin{figure}[h]
    \centering
    \includegraphics[width=\textwidth]{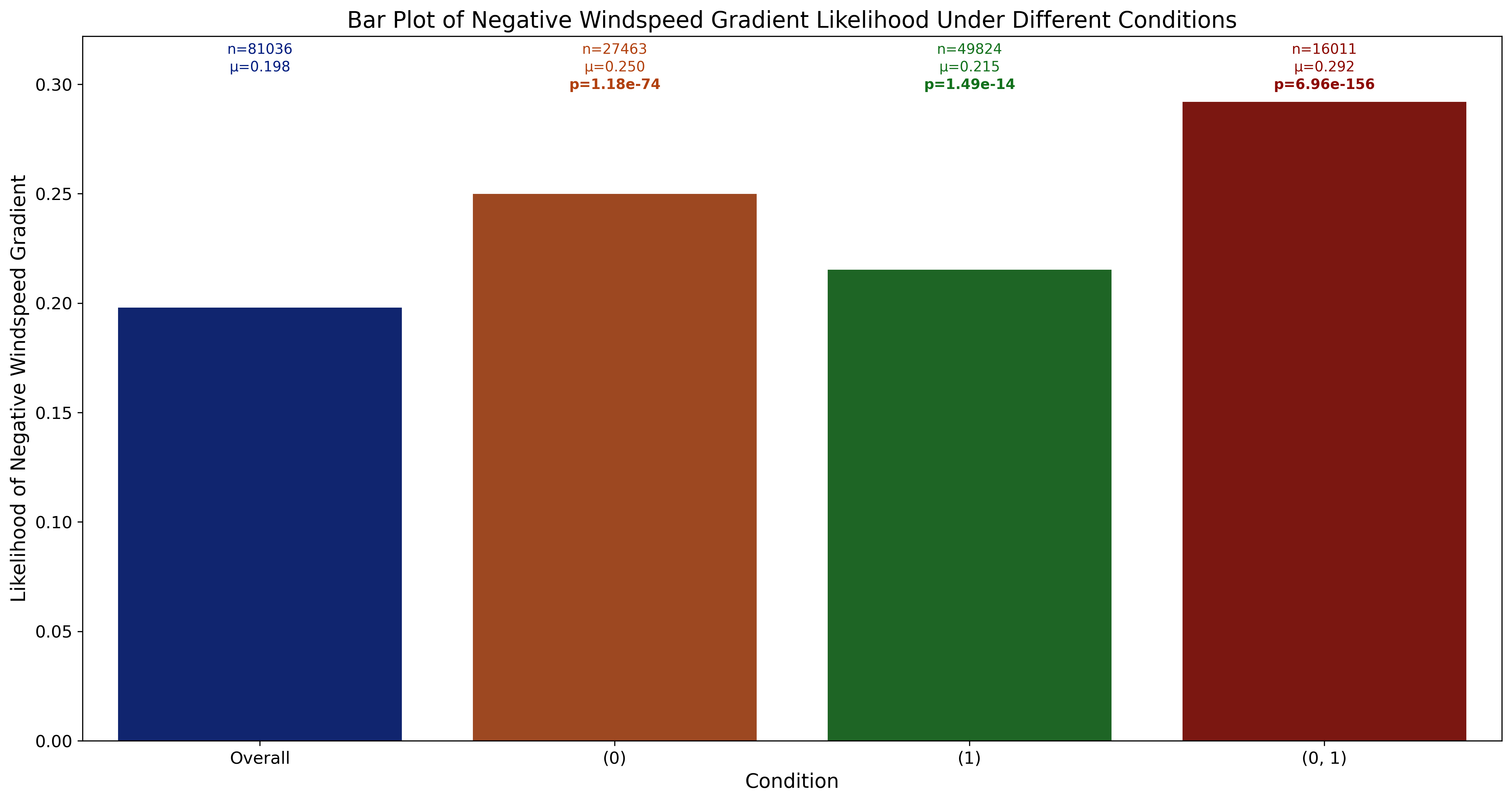}
        \caption{Bar plot showing the likelihood of negative wind speed gradient under the conditions below. $n$ is the number of observations available; $\mu$ is the proportion of these observations with negative wind speed gradient; $p$ is the p-value from a z-test for whether this proportion is 
        higher than in the Overall set.
    \\ \hspace*{2em}(0) \texttt{wind\_direction}: from land
    \\ \hspace*{2em}(1) $0$km $ \leq \texttt{distance\_from\_coast} \leq 3.18$km
    }
\label{fig:discovery-1}
\end{figure}

The Discovery Engine identified a strong relationship between wind direction and the likelihood of negative wind speed gradients. Specifically, gradients that contradict the expected monotonic increase in wind speed with height within the surface layer were substantially more frequent under offshore (from-land) wind conditions, especially when close to the coast. The bar plot in Figure~\ref{fig:discovery-1} visualises this difference. It shows that being close to land and wind coming from land are separately and jointly associated with a statistically significant increase in the likelihood of negative wind speed gradients. 

This finding serves as the first indication that negative wind speed gradients are not randomly distributed, but instead follow systematic patterns tied to environmental context. To further stress test this result, which The Discovery Engine autonomously uncovered, we now examine how their prevalence varies with coastal proximity, wind direction, and their co-occurrence with extreme thermodynamic gradients.

\subsection{Investigating the Effect of Offshore Winds}

\subsubsection{Prevalence of Negative Wind Speed Gradients Across Campaign Phases}

\begin{figure}[h]
    \centering
    \includegraphics[width=1\textwidth]{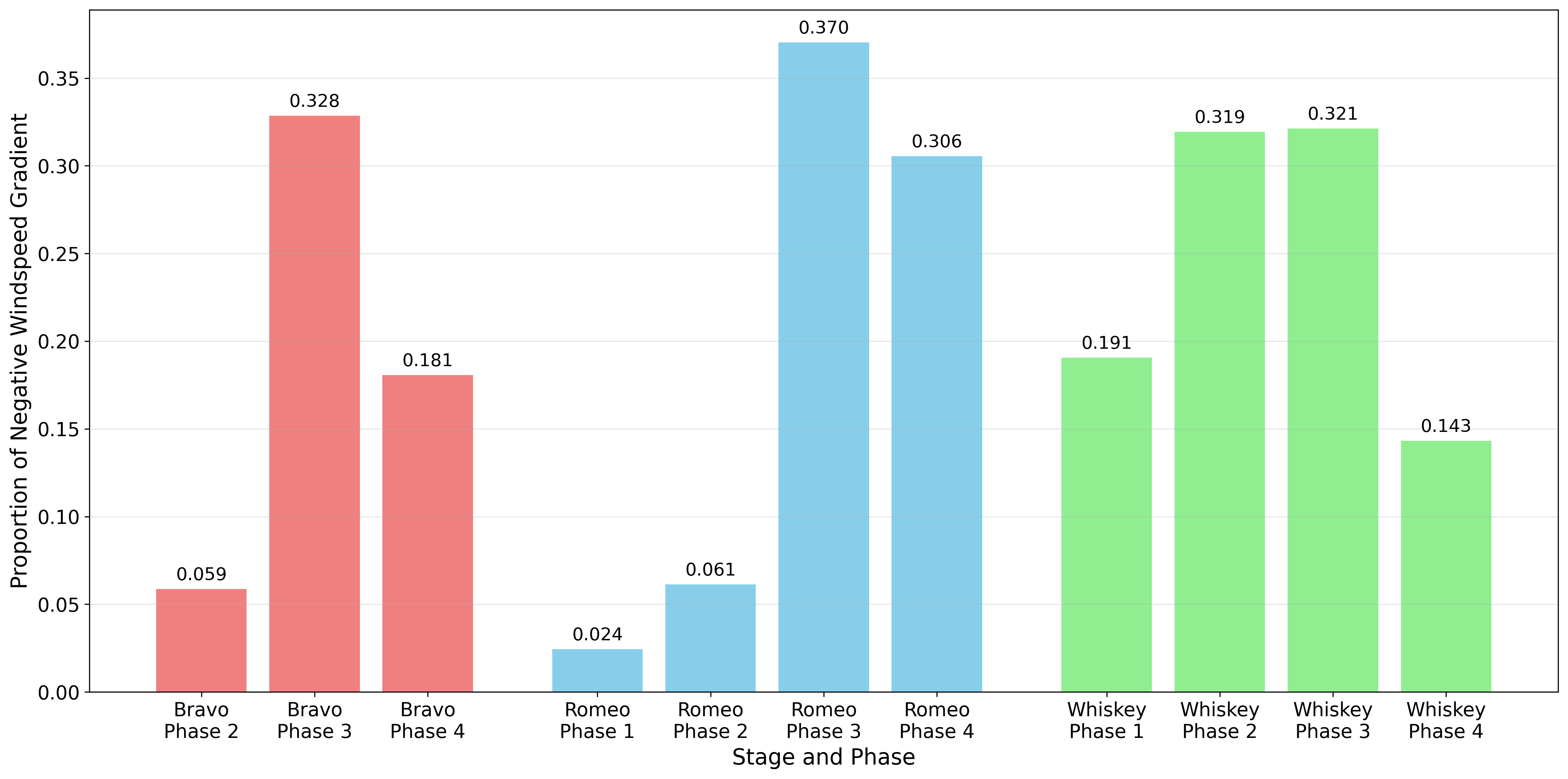}
    \caption{Proportion of observations with negative wind speed gradient, by stage and phase of study.}
    \label{fig:neg_gradient}
\end{figure}

Negative wind speed gradients occurred in 19.8\% of observations across the combined dataset, though this rate varied across different phases of the campaign. Figure~\ref{fig:neg_gradient} presents the frequency of negative wind speed gradients stratified by study phase, revealing significant variation across phases.

\subsubsection{Effects of Coastal Proximity and Wind Direction}

The probability of observing a negative wind speed gradient was strongly modulated by two key factors: the buoy's distance from land, and the direction of the prevailing wind. We categorised observations by whether the wind was blowing from land or from sea, and by proximity to shore. 

\begin{figure}[h]
    \centering
    \includegraphics[width=\textwidth]{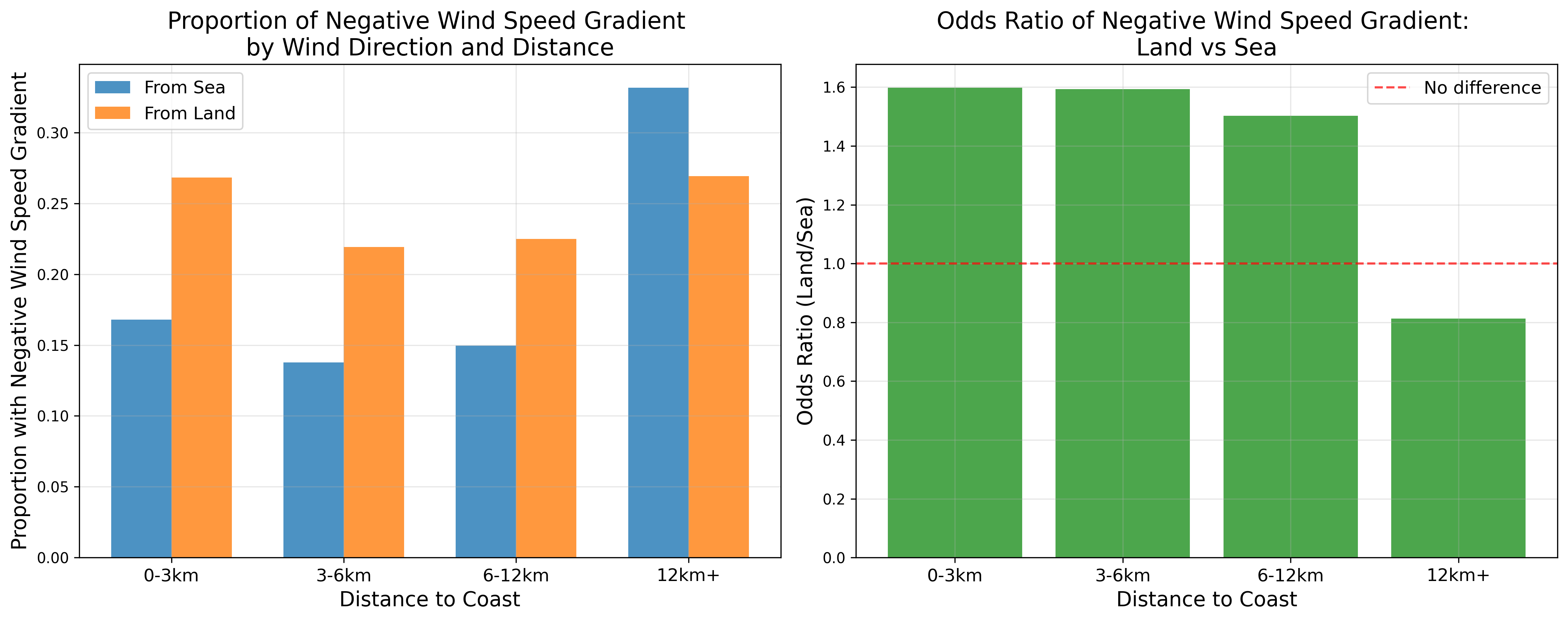}
    \caption{Left: Proportion of observations with negative wind speed gradient, by wind direction (from land or sea) and buoy distance to land. Right: Odds ratios of negative wind speed gradient, comparing when wind direction is from land vs from sea.}
    \label{fig:neg_gradient_odds}
\end{figure}

Figure~\ref{fig:neg_gradient_odds} shows that negative gradients were significantly more likely when the wind blew offshore, particularly for the buoys nearest to land. For the buoys more than 12km from land, the opposite relationship holds, indicating that the distorting effects of coastal winds diminish with distance from the coast. We quantify this using odds ratios, which demonstrate a clear interaction between wind direction and distance from land in their relationship with negative wind speed gradients. The values presented in Figure~\ref{fig:neg_gradient_odds} are displayed in Table~\ref{tab:wind-distance} along with their associated p-values, which show extremely significant relationships for those buoys within 12km of land.

\begin{table}[h]
    \centering
    \begin{tabular}{llcccc}
        \toprule
        & & \multicolumn{4}{c}{\textbf{Distance from Land}} \\
        \cmidrule(lr){3-6}
        & & 0--3 km & 3--6 km & 6--12 km & 12+ km \\
        \midrule
        \multirow{2}{*}{\textbf{Wind Direction}} 
        & From Sea & 0.168 & 0.138 & 0.150 & 0.332 \\
        & From Land  & 0.268 & 0.219 & 0.225 & 0.270 \\
        \addlinespace
        \multicolumn{2}{l}{\textbf{Odds Ratio (Land / Sea)}} & 1.60 & 1.59 & 1.50 & 0.81 \\
        \addlinespace
        \multicolumn{2}{l}{\textbf{P-value}} & $6.1\times 10^{-130}$ & $2.2 \times 10^{-55}$ & $1.8 \times 10^{-17}$ & $\approx 1$ \\
        \bottomrule
    \end{tabular}
    \caption{Values and odds ratios presented in Figure~\ref{fig:neg_gradient_odds}. Odds ratios compare from-land to from-sea conditions at each distance.}
    \label{tab:wind-distance}
\end{table}


\subsubsection{Strong Flux Gradients and Departures from MOST Assumptions}

\begin{figure}[p]
    \centering
    \includegraphics[width=\textwidth]{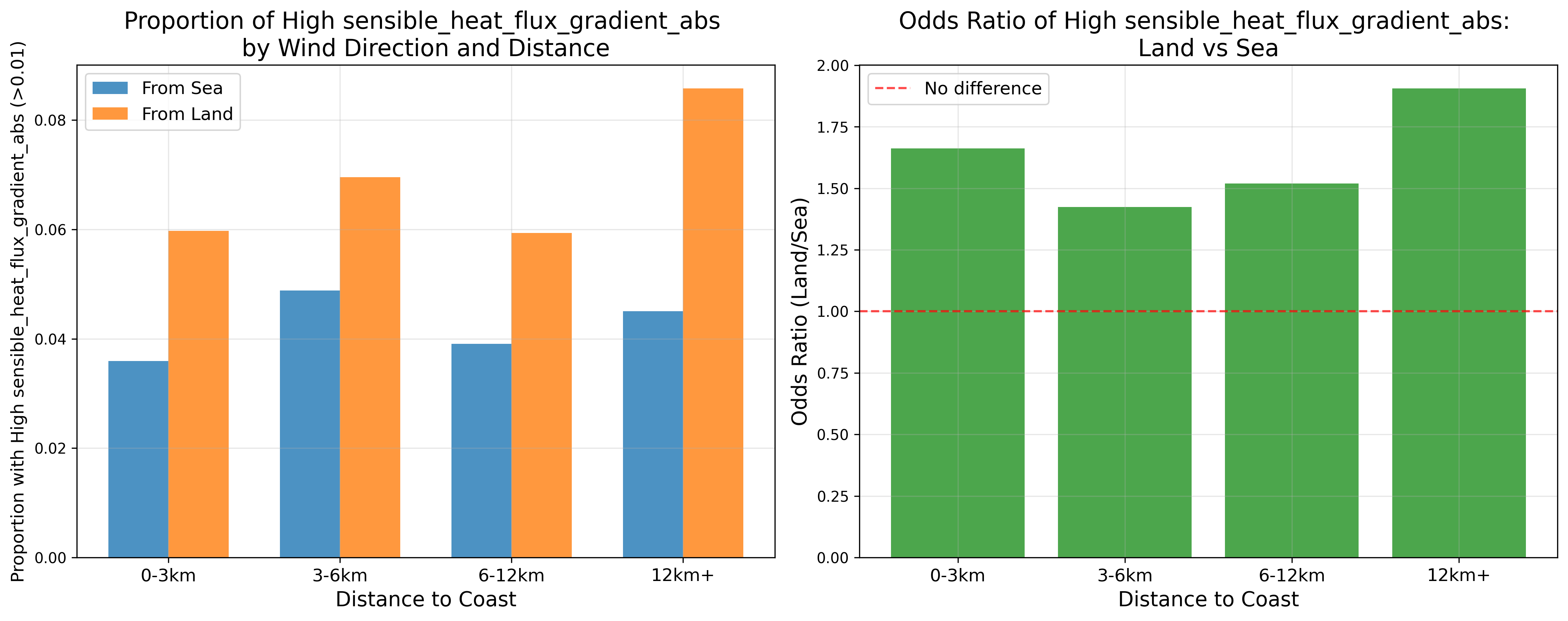}
    \caption{Left: Proportion of observations with absolute sensible heat flux gradient greater than 0.01 $K \cdot s^{-1}$, by wind direction (from land or sea) and buoy distance to land. Right: Odds ratios of these outcomes, comparing when wind direction is from land vs from sea.}
    \label{fig:flux_gradient_odds}
\end{figure}

We also observed a marked increase in the frequency of strong vertical gradients in sensible heat flux under offshore wind conditions at all distances from the coast. Figure~\ref{fig:flux_gradient_odds} presents this result in the same stratified format as the previous figure, highlighting that both types of departures from MOST assumptions (kinematic and thermodynamic) share a common dependence on wind direction and coastal proximity. The values presented in Figure~\ref{fig:flux_gradient_odds} are displayed in Table~\ref{tab:flux-extreme} along with their associated p-values, which show significant relationships for buoys at all distances. 





\begin{table}[p]
    \centering
    \begin{tabular}{llcccc}
        \toprule
        & & \multicolumn{4}{c}{\textbf{Distance from Land}} \\
        \cmidrule(lr){3-6}
        & & 0--3 km & 3--6 km & 6--12 km & 12+ km \\
        \midrule
        \multirow{2}{*}{\textbf{Wind Direction}} 
        & From Sea & 0.036 & 0.049 & 0.039 & 0.045 \\
        & From Land  & 0.060 & 0.070 & 0.059 & 0.086 \\
        \addlinespace
        \multicolumn{2}{l}{\textbf{Odds Ratio (Land / Sea)}} & 1.66 & 1.42 & 1.51 & 1.90 \\
        \addlinespace
        \multicolumn{2}{l}{\textbf{P-value}} & $1.1\times 10^{-29}$ & $7.0\times 10^{-11}$ & $2.0\times 10^{-5}$ & $2.5\times 10^{-13}$ \\
        \bottomrule
    \end{tabular}
    \caption{Large ($>0.01 K \cdot s^{-1}$) absolute sensible heat flux gradient probability by wind direction and buoy distance from land. Odds ratios compare from-land to from-sea conditions at each distance.}
    \label{tab:flux-extreme}
\end{table}



\subsubsection{Co-occurrence of Kinematic and Thermodynamic Anomalies}

To assess whether the two anomalies tend to co-occur, we tested for statistical association between negative wind speed gradients and extreme sensible heat flux gradients. As shown in Table~\ref{tab:cooccurrence}, the two phenomena are strongly linked: observations with negative wind speed gradients are over 2.5 times more likely to exhibit extreme flux gradients than those with positive gradients. This association is highly statistically significant and suggests that both anomalies may arise from a shared set of dynamical conditions.

\begin{table}[p]
\centering
\caption{
Negative wind speed gradients are strongly associated with extreme ($>0.01 K \cdot s^{-1}$) vertical gradients in sensible heat flux. The 2×2 table below shows co-occurrence counts ($n = 81{,}036$) and the result of a one-sided Fisher’s exact test for positive association. Observations with negative wind speed gradients were 2.54× more likely to exhibit extreme SHFG than those with positive gradients, a statistically significant relationship.
}
\label{tab:cooccurrence}

\vspace{0.5em}
\begin{tabular}{lcc}
\toprule
& \textbf{Extreme SHFG} & \textbf{Normal SHFG} \\
\midrule
\textbf{Wind Speed Gradient $<$ 0} & 1{,}476 & 14{,}560 \\
\textbf{Wind Speed Gradient $\geq$ 0} & 2{,}494 & 62{,}506 \\
\bottomrule
\end{tabular}

\vspace{1em}
\small
Odds ratio: 2.54 \\
95\% Confidence Interval: [2.38, 2.72] \\
One-sided Fisher’s exact test p-value: $2.1\times 10^{-50}$

\end{table}

\subsubsection{Interpretation in Terms of Coastal Internal Boundary Layers}

Taken together, these results provide strong empirical evidence for the hypothesis described in Section~\ref{sec:internal-boundary-layers} that coastal internal boundary layers, which represent a specific nearshore case of thermally stable boundary layers, are a dominant contributor to the observed anomalies. The consistent modulation of anomalies by wind direction and distance to shore suggests the influence of coastal adjustment processes.

\subsection{Conditions Most Strongly Associated with MOST Departures Near Shore under Offshore Winds}
To better understand the environmental context of MOST departures in nearshore settings, we used the Discovery Engine to identify the conditions most strongly associated with anomalies during offshore wind flow. Focusing on observations close to shore with wind blowing from land, it extracted the most distinctive combinations of features associated with both negative wind speed gradients and extreme vertical gradients in sensible heat flux. 

\begin{figure}[h]
    \centering
    \includegraphics[width=0.9\textwidth]{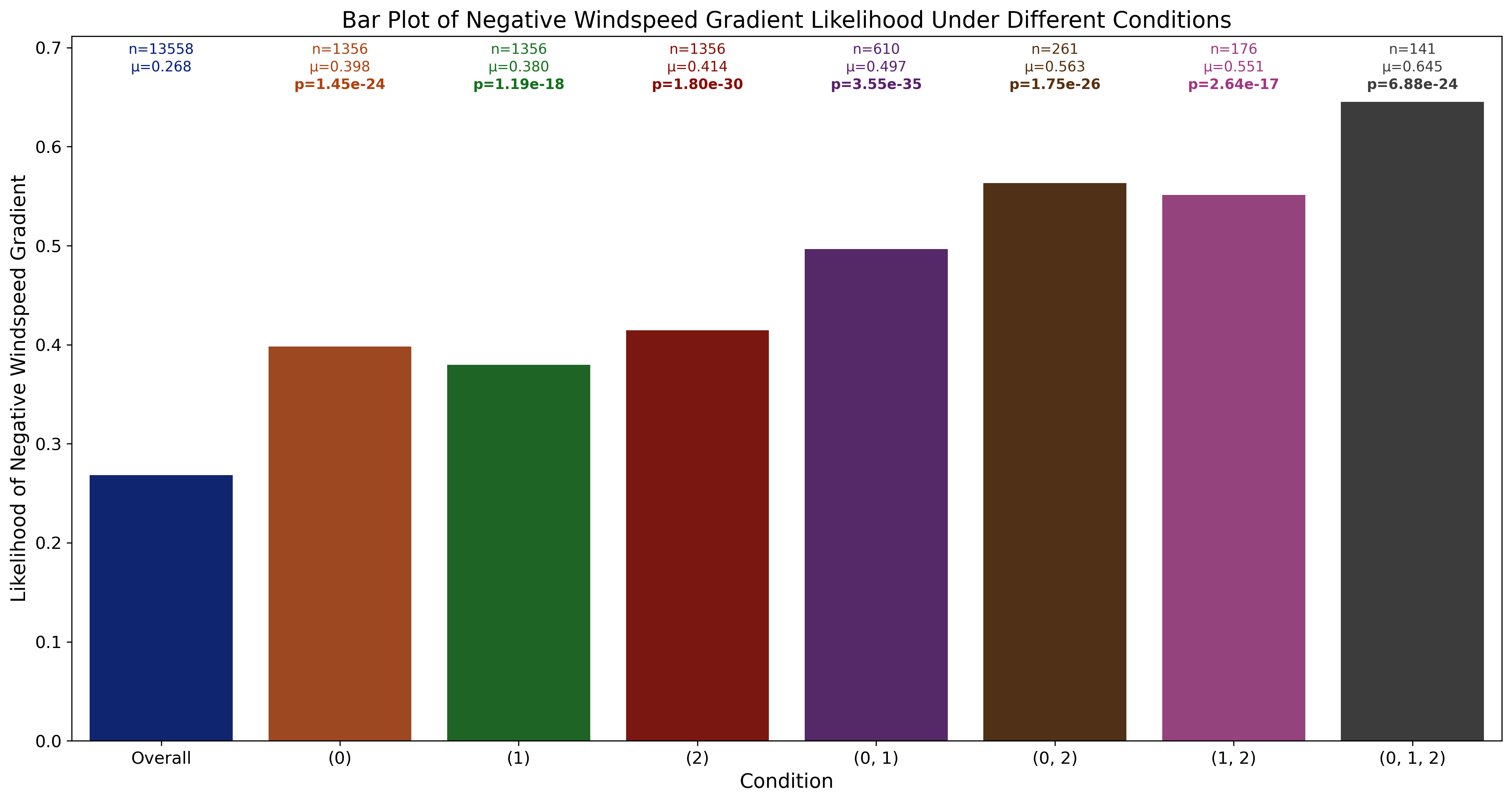}
    \caption{Bar plot showing the likelihood of negative wind speed gradient under the conditions below, with data restricted to observations within 3km of the coast, with offshore winds. $n$ is the number of observations available; $\mu$ is the proportion of these observations with negative wind speed gradient; $p$ is the p-value from a z-test for whether this proportion is higher than in the Overall set.
    \\ \hspace*{2em}(0) $98.8 \leq \texttt{RH} \leq 99.7 \quad (90\%\text{ to }100\%\text{ quantile})$
    \\ \hspace*{2em}(1) $283.9 \leq \texttt{T} \leq 285.8 \quad (0\%\text{ to }10\%\text{ quantile})$
    \\ \hspace*{2em}(2) $995.8 \leq \texttt{P} \leq 1004.3 \quad (0\%\text{ to }10\%\text{ quantile})$
    }
    \label{fig:wspd-further-pattern-1}
\end{figure}

\begin{figure}[h]
    \centering
    \includegraphics[width=0.9\textwidth]{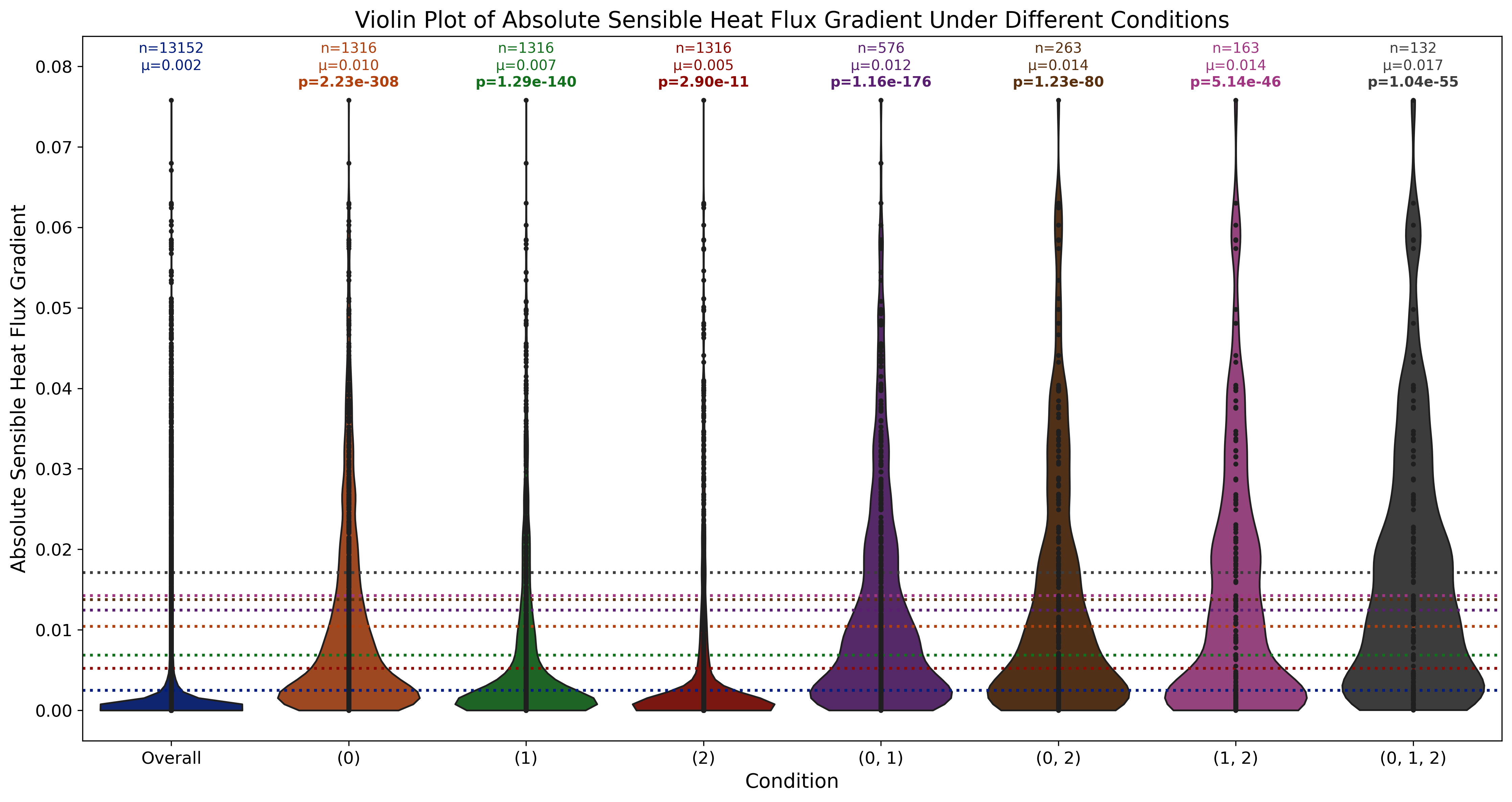}
    \caption{Violin plot showing the distribution of absolute \texttt{sensible\_heat\_flux\_gradient} under the conditions below, with data restricted to observations within 3km of the coast, with offshore winds. $n$ is the number of observations available; $\mu$ is the mean of these observations; $p$ is the p-value from a Mann-Whitney for whether the distribution of these observations is higher than in the Overall set.
    \\ \hspace*{2em}(0) $98.8 \leq \texttt{RH} \leq 99.7 \quad (90\%\text{ to }100\%\text{ quantile})$
    \\ \hspace*{2em}(1) $283.9 \leq \texttt{T} \leq 285.8 \quad (0\%\text{ to }10\%\text{ quantile})$
    \\ \hspace*{2em}(2) $995.8 \leq \texttt{P} \leq 1004.3 \quad (0\%\text{ to }10\%\text{ quantile})$
    }
    \label{fig:shfg-further-pattern-1}
\end{figure}

This approach revealed a small set of physical conditions: low air temperature, low pressure, and high relative humidity. These conditions consistently co-occur with the strongest departures from MOST, namely, negative wind speed gradients and extreme sensible heat flux gradients. Figures~\ref{fig:wspd-further-pattern-1} and~\ref{fig:shfg-further-pattern-1} illustrate how the likelihood of these anomalies increases as more of these conditions are jointly satisfied.

The combination of low air temperature, low pressure, and high relative humidity near the surface suggests that the offshore flow is rapidly adjusting to the cooler, moister marine environment. This adjustment likely leads to the development of a shallow stable layer near the surface, which suppresses turbulent mixing and produces sharp vertical gradients in sensible heat flux. While these conditions are consistent with the formation of an internal boundary layer (Section~\ref{sec:internal-boundary-layers}), further targeted observations would be needed to directly confirm this mechanism as the driver of the anomalies observed near the coast.

\subsection{Further Discoveries Far From Shore}
While the most frequent departures from MOST occurred near the coast under offshore winds, we also applied the Discovery Engine to observations farther from shore to explore whether different patterns might emerge in open-ocean settings. This analysis highlighted two distinct regimes, each consistent with a separate hypothesis from the literature: wave-driven wind jets (Section~\ref{sec:wave-driven-wind-jets}) and thermally stable boundary layers (Section~\ref{sec:internal-boundary-layers}). These findings indicate that different physical processes may contribute to MOST departures in offshore environments compared to coastal zones.

\subsubsection{Support for \emph{Wave-Driven Wind Jets} Far From Shore}

In offshore environments far from land, the Discovery Engine identified a distinct regime in which negative wind speed gradients became more frequent. This regime was characterised by high wave age and low wind speed—conditions known to decouple the near-surface atmosphere from local turbulent production and amplify the influence of wave-driven processes. As detailed in Section~\ref{sec:wave-driven-wind-jets}, swell-generated waves traveling faster than the overlying wind can transfer momentum upward into the atmosphere, forming low-level wind maxima inconsistent with MOST’s assumptions of monotonic vertical profiles.

\begin{figure}[p]
    \centering
    \includegraphics[width=0.9\textwidth]{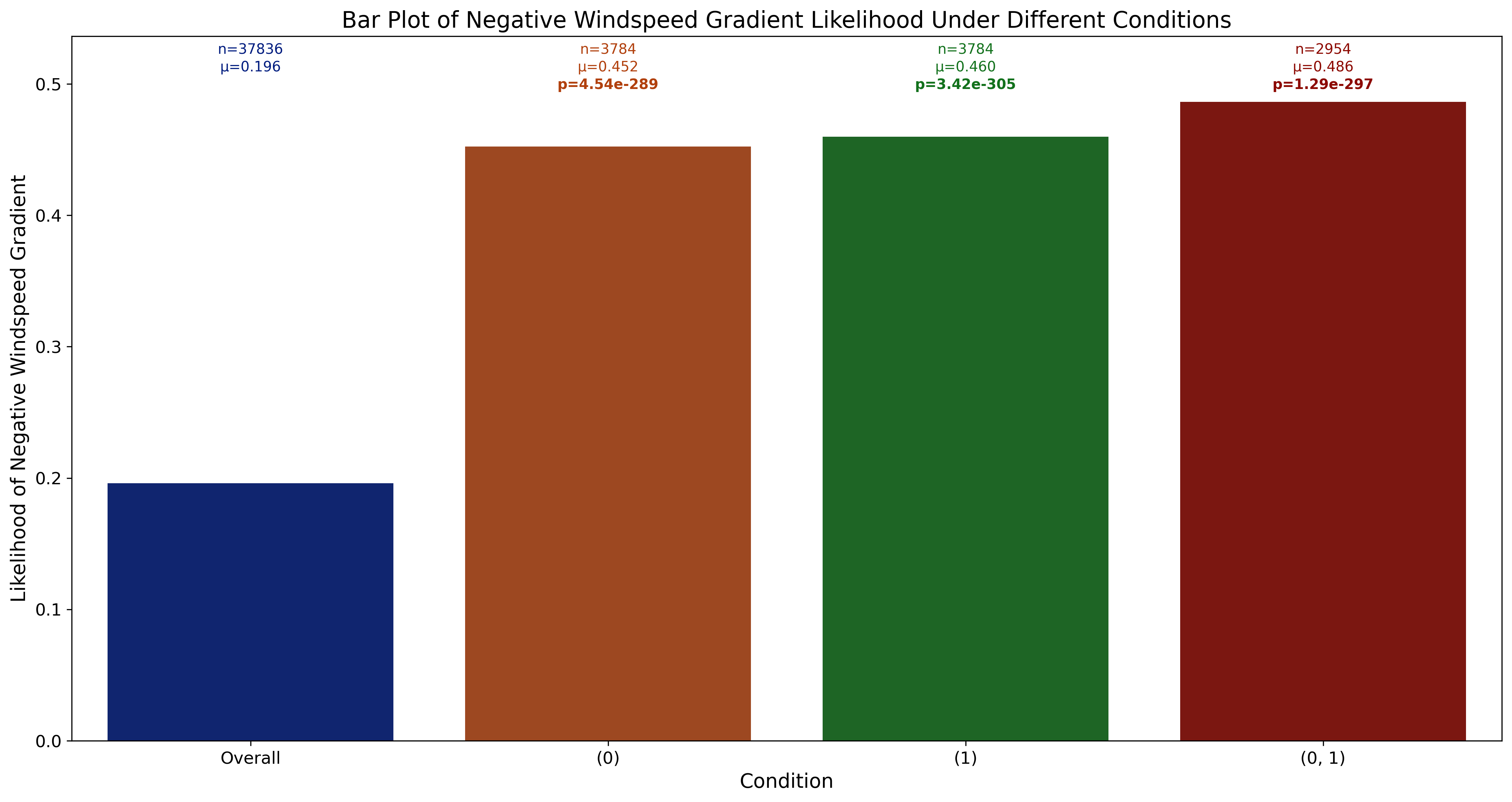}
    \caption{Bar plot showing the likelihood of negative wind speed gradient under the conditions below, with data restricted to observations greater than 3km from the coast. $n$ is the number of observations available; $\mu$ is the proportion of these observations with negative wind speed gradient; $p$ is the p-value from a z-test for whether this proportion is higher than in the Overall set.
    \\ \hspace*{2em}(0) $5.28 \leq \texttt{wave\_age} \leq 44.41 \quad (90\%\text{ to }100\%\text{ quantile})$
    \\ \hspace*{2em}(1) $0.17 \leq \texttt{wspd} \leq 1.78 \quad (0\%\text{ to }10\%\text{ quantile})$
    }
    \label{fig:wspd-further-pattern-2}
\end{figure}

Figure~\ref{fig:wspd-further-pattern-2} shows that the combination of high wave age (wave phase speed greatly exceeding local wind speed) and weak winds is strongly associated with the occurrence of negative wind speed gradients. These findings are consistent with prior observations of swell-induced wind jets, in which the vertical profile of wind speed reflects the imprint of wave-supported momentum transfer rather than classical shear turbulence. Importantly, the persistence of this signal far from the coast suggests that wave–atmosphere coupling may play a dominant role in shaping boundary-layer structure over open water, independent of coastal geometry.

\subsubsection{Support for \emph{Thermal Stratification and Stable Layers} Far From Shore}
A distinct regime associated with extreme vertical gradients in sensible heat flux was identified by the Discovery Engine far from shore. This regime is characterised by high relative humidity, negative sensible heat flux, and large positive air–sea temperature differences (i.e., the air significantly warmer than the sea surface). Together, these conditions point to strong thermal stratification: the warm, moist air overlying cooler water creates a stable boundary layer in which buoyancy suppresses turbulent mixing, leading to sharp vertical gradients in heat flux.

Unlike the coastal internal boundary layers described earlier, this thermally stable boundary layer forms in the open ocean, driven by persistent local air–sea contrasts rather than land–sea transitions. The presence of negative sensible heat flux under these conditions confirms that heat is being transferred downward into the ocean, further reinforcing stability. The added predictive power of the air–sea temperature difference condition highlights the direct link between the observed anomalies and their physical cause: buoyancy suppression of turbulence due to thermal stratification.

\begin{figure}[p]
    \centering
    \includegraphics[width=0.9\textwidth]{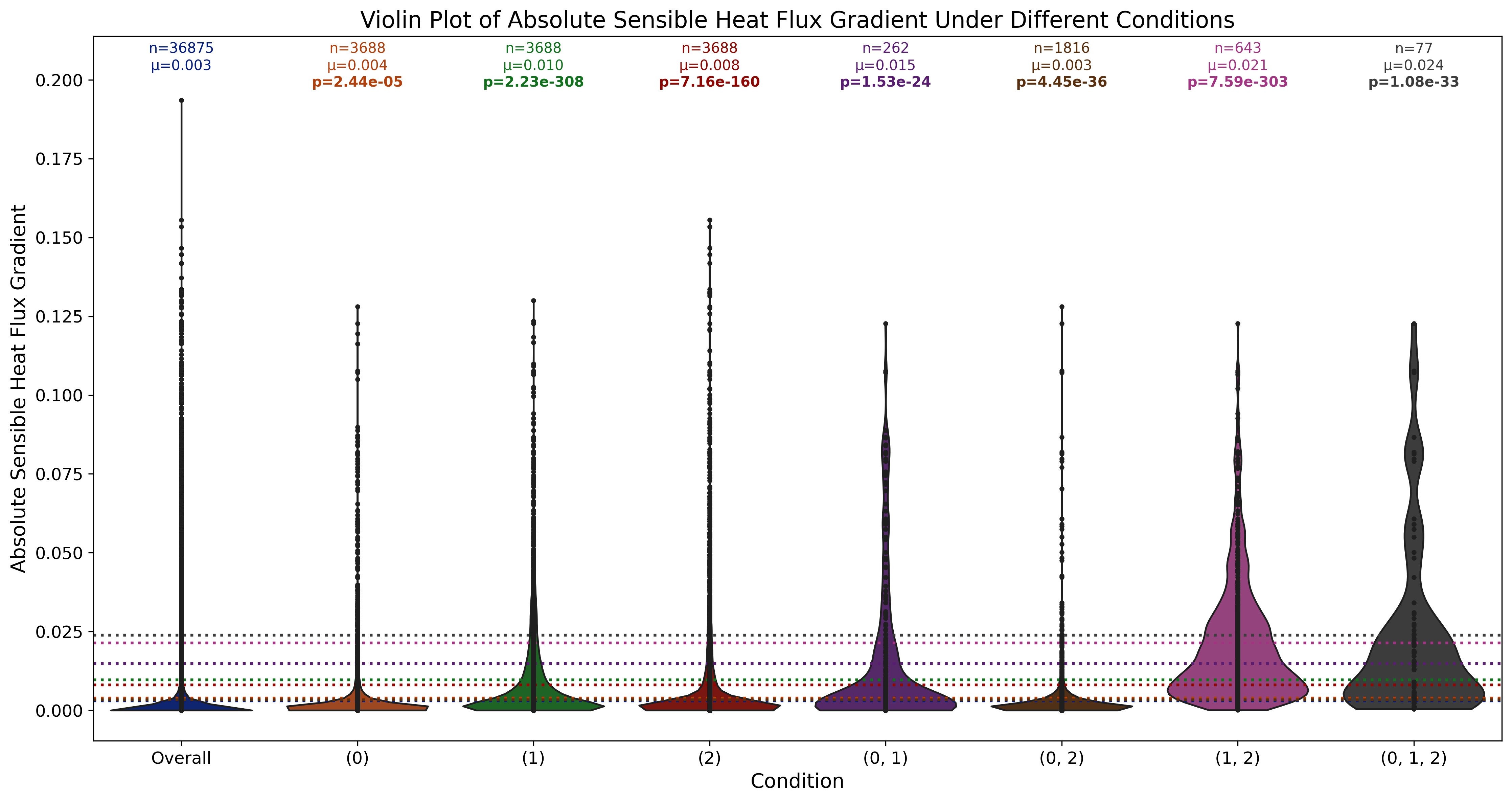}
    \caption{Violin plot showing the distribution of absolute \texttt{sensible\_heat\_flux\_gradient} under the conditions below, with data restricted to observations greater than 3km from the coast. $n$ is the number of observations available; $\mu$ is the mean of these observations; $p$ is the p-value from a Mann-Whitney for whether the distribution of these observations is higher than in the Overall set.
    \\ \hspace*{2em}(0) $1.05 \leq \texttt{T\_gradient} \leq 3.95 \quad (90\%\text{ to }100\%\text{ quantile})$
    \\ \hspace*{2em}(1) $97.8 \leq \texttt{RH} \leq 100.1 \quad (90\%\text{ to }100\%\text{ quantile})$
    \\ \hspace*{2em}(2) $-0.16 \leq \texttt{sensible\_heat\_flux} \leq -0.01 \quad (0\%\text{ to }10\%\text{ quantile})$
    }
    \label{fig:shfg-further-pattern-2}
\end{figure}

These results align strongly with the stable-layer hypothesis presented in Section~\ref{sec:thermal-stratification}. As shown in Figure~\ref{fig:shfg-further-pattern-2}, the likelihood of a large sensible heat flux gradient increases substantially under these conditions, providing robust empirical support for the role of thermal stratification in driving surface-layer anomalies even far from land.

\section{Interpretation and Discussion}\label{discussion}

\subsection{MOST Assumptions Often Do Not Hold Near the Sea Surface}

Our findings reveal that the assumptions of MOST do not hold under a wide range of real-world coastal and marine conditions, within 10~m of the sea surface. In particular, we observe frequent instances where wind speed does not increase with height as expected under MOST assumptions, along with large vertical variations in sensible heat flux. These two anomalies occur together more often than expected by chance and follow structured patterns related to environmental context. This suggests that conditions leading to the breakdown of MOST assumptions are not random or due to measurement noise, but arise from identifiable physical mechanisms that can now be mapped and predicted.

\subsection{Explaining the Patterns with Known Mechanisms}

The structure in the observed anomalies aligns with several well-established mechanisms known to produce departures from MOST assumptions. The clearest support is found for coastal internal boundary layers. These form when air flows from land over the ocean and adjusts to the sharp change in surface roughness and thermal properties, producing layered structures with strong vertical transitions. Under these conditions, we observe the most frequent joint occurrence of negative wind speed gradients and extreme sensible heat flux gradients. These patterns are most pronounced near the coast under offshore wind flow, where the development of a stable near-surface layer suppresses mixing and leads to sharp vertical gradients.

Further from the coast, two additional mechanisms emerge. One regime is characterised by high wave age and low wind speeds. This combination supports the formation of wave-driven wind jets, in which swell transfers momentum upward into the atmosphere, creating non-monotonic wind profiles. A second regime involves high relative humidity, negative sensible heat flux, and large positive air–sea temperature differences, favouring thermally stable boundary layers with suppressed turbulence and sharp vertical flux gradients. These offshore regimes are physically distinct from the coastal ones and highlight the diversity of processes that can lead to departures from MOST.

\subsection{Discovery Engine as a Driver of Scientific Insight}

The Discovery Engine played a central role in revealing these patterns. It did not simply fit a model; it extracted precise, interpretable combinations of environmental conditions that strongly predict theory-breaking behaviour. In doing so, it surfaced evidence for existing hypotheses. This allowed us to go beyond simple correlations and identify causal candidates for surface layer disruption. By scanning a large space of possible feature interactions and validating them against the full dataset, the system avoided human bias while still enabling expert interpretation. 

\subsection{Implications for Air–Sea Flux Modelling}


The prevalence and structure of departures from MOST assumptions have direct implications for surface flux estimation. MOST-based bulk flux algorithms, such as COARE \citep{fairall2003bulk}, the GFS surface flux scheme \citep{ncep2003gfs}, and WRF flux parameterisations \citep{wrf2008, wrf2019}, are designed to account for stability effects and, in some cases, wave influences. However, small-scale features such as internal boundary layers and wave-induced low-level jets can arise at vertical scales or with variability that remains difficult to resolve within these frameworks. These effects may introduce systematic flux errors during offshore wind events near the coast. Offshore, subtle interactions between wave fields and stratification may also contribute to underestimated variability. The precise implications for modelling remain an open question, warranting further investigation.

Our results suggest that future flux parameterisations could benefit from incorporating simple conditional checks — for example, flags for offshore wind, proximity to land, or high wave age — to help identify regimes where MOST assumptions are more likely to break down. While such checks would not capture all cases of surface-layer anomalies, they could provide a practical first step towards improving flux estimates by signalling conditions where alternative parameterisations or enhanced modelling strategies may be needed.

\subsection{Limitations and Future Directions}

This study is limited to specific deployments in particular coastal regions. While the patterns appear robust across platforms and phases, they should be tested in other locations and seasons. In addition, although the Discovery Engine produces interpretable insights, its outputs should be treated as hypotheses requiring physical validation, especially when extrapolated to new contexts.

\section{Conclusion}\label{sec:conclusion}

We have shown that departures from MOST assumptions are common, predictable, and physically structured. Using the Discovery Engine, we uncovered clear environmental regimes, especially offshore wind events near the coast, that are strongly associated with non-monotonic wind profiles and large vertical gradients in sensible heat flux. These anomalies align with known mechanisms such as internal boundary layers, wave-driven jets, and thermal stratification driven by large air–sea temperature differences, depending on the setting.

This data-driven approach demonstrates a new way of identifying and explaining surprising patterns in atmospheric data. By combining large-scale environmental data with interpretable machine learning, we move closer to building more accurate and context-aware models of air–sea interaction. The patterns uncovered here offer a foundation for future parameterisation schemes that are both physically informed and practically deployable.

\section{Acknowledgements}
The authors would like to thank Prof. Milan Curcic of the University of Miami for his guidance and providing data from the CLASI field campaign. PH and SEH are supported by the NSF National Center for Atmospheric Research, which is a major facility sponsored by the U.S. National Science Foundation under Cooperative Agreement No. 1852977. Funding for PH and SEH was provided by the DOE-funded ORACLE project as a subcontractor to Pacific Northwest National Laboratory under award \#778383.

\newpage
\bibliographystyle{plainnat}
\bibliography{refs.bib}

\newpage
\appendix
\section{Feature Distribution Plots}
\begin{figure}[htbp]
    \centering

    \begin{subfigure}[b]{0.35\textwidth}
        \includegraphics[width=\textwidth]{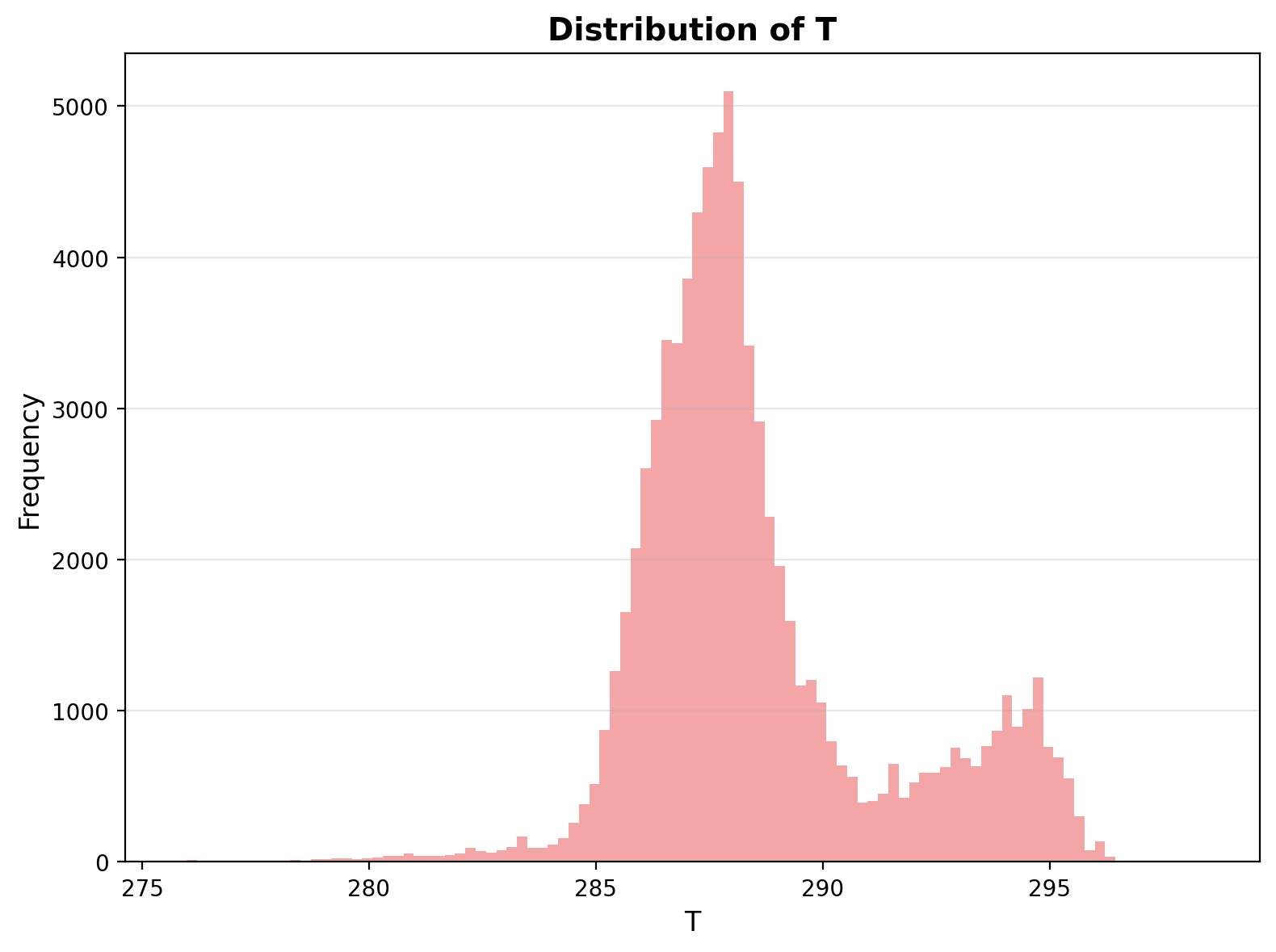}
        \caption{Histogram of values of \texttt{T} (temperature) in Kelvin.}
    \end{subfigure}
    \hfill
    \begin{subfigure}[b]{0.35\textwidth}
        \includegraphics[width=\textwidth]{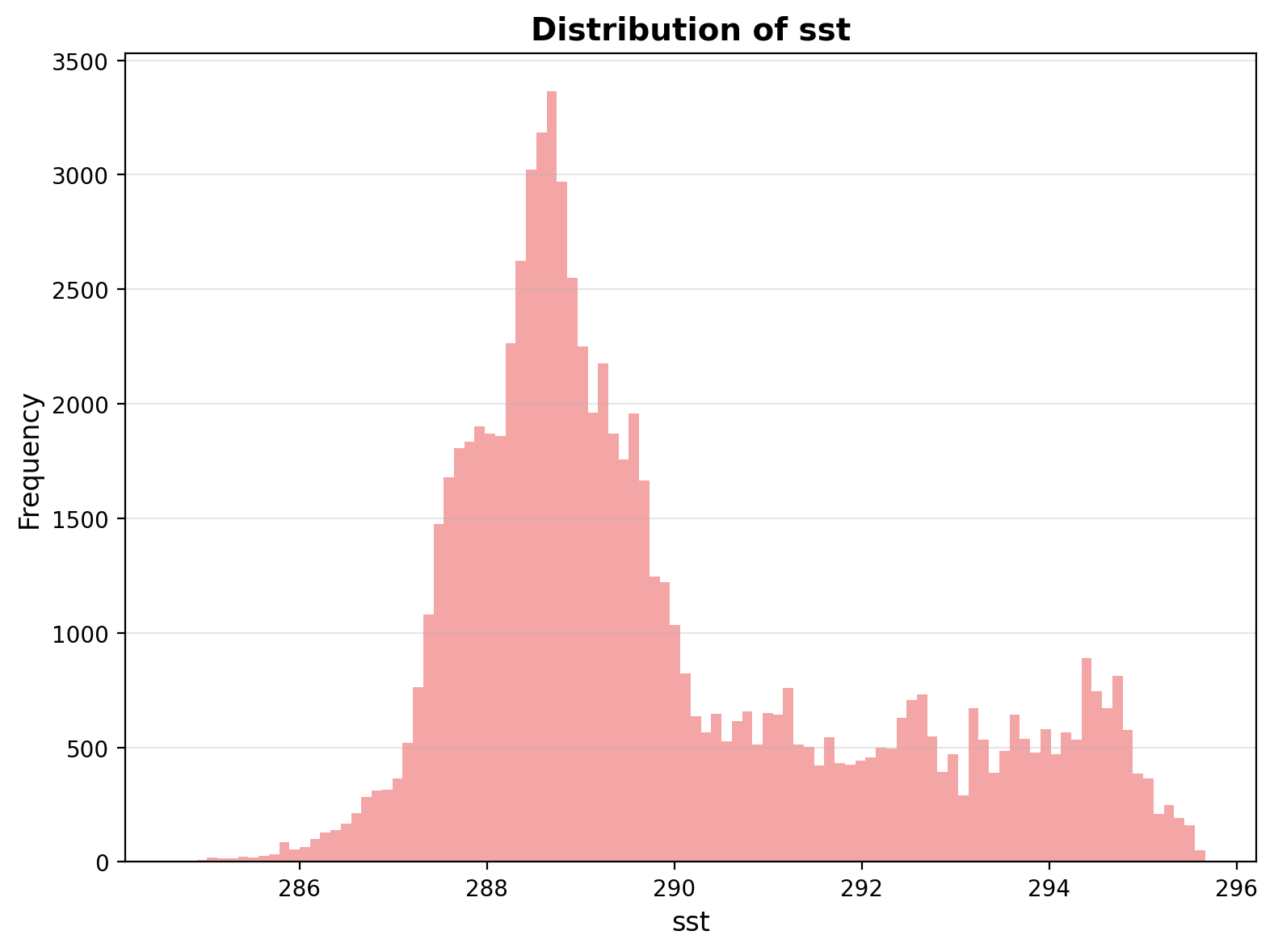}
        \caption{Histogram of values of \texttt{sst} (sea-surface temperature) in Kelvin.}
    \end{subfigure}

    \vspace{1em}

    \begin{subfigure}[b]{0.35\textwidth}
        \includegraphics[width=\textwidth]{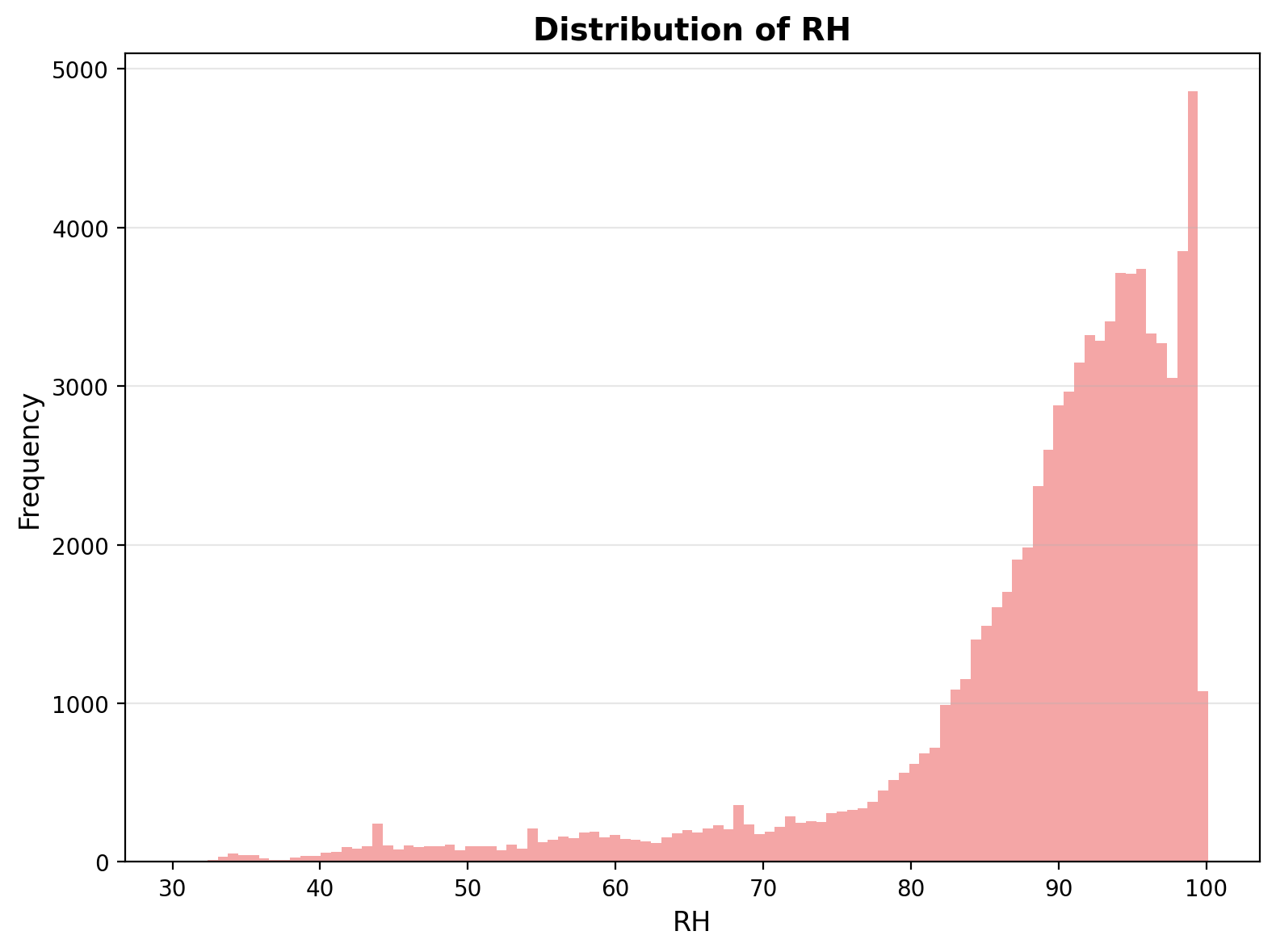}
        \caption{Histogram of values of \texttt{RH} (relative humidity) in \%.}
    \end{subfigure}
    \hfill
    \begin{subfigure}[b]{0.35\textwidth}
        \includegraphics[width=\textwidth]{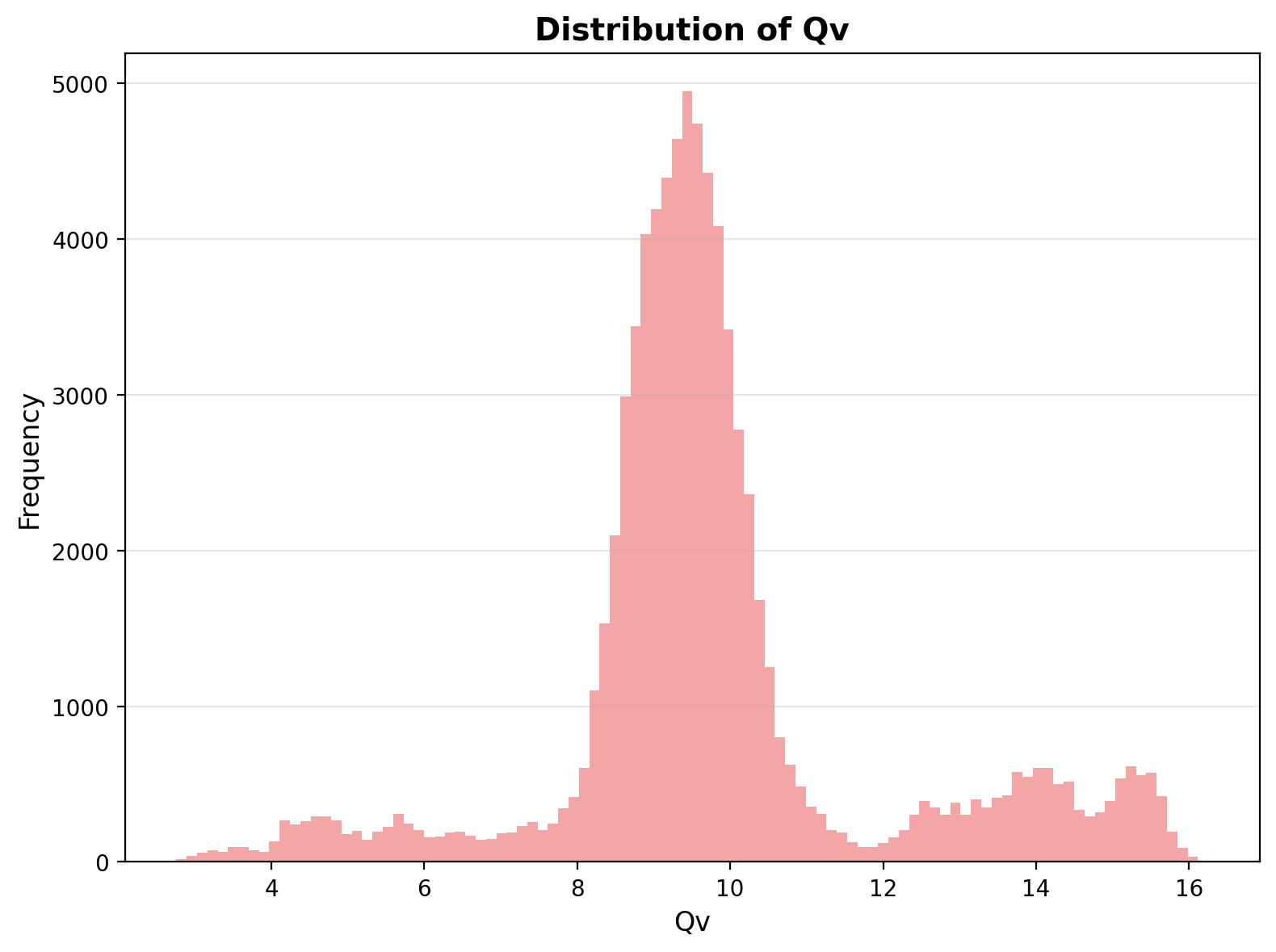}
        \caption{Histogram of values of \texttt{Qv} (water vapour mixing ratio) in g/Kg.}
    \end{subfigure}

    \vspace{1em}

    \begin{subfigure}[b]{0.35\textwidth}
        \includegraphics[width=\textwidth]{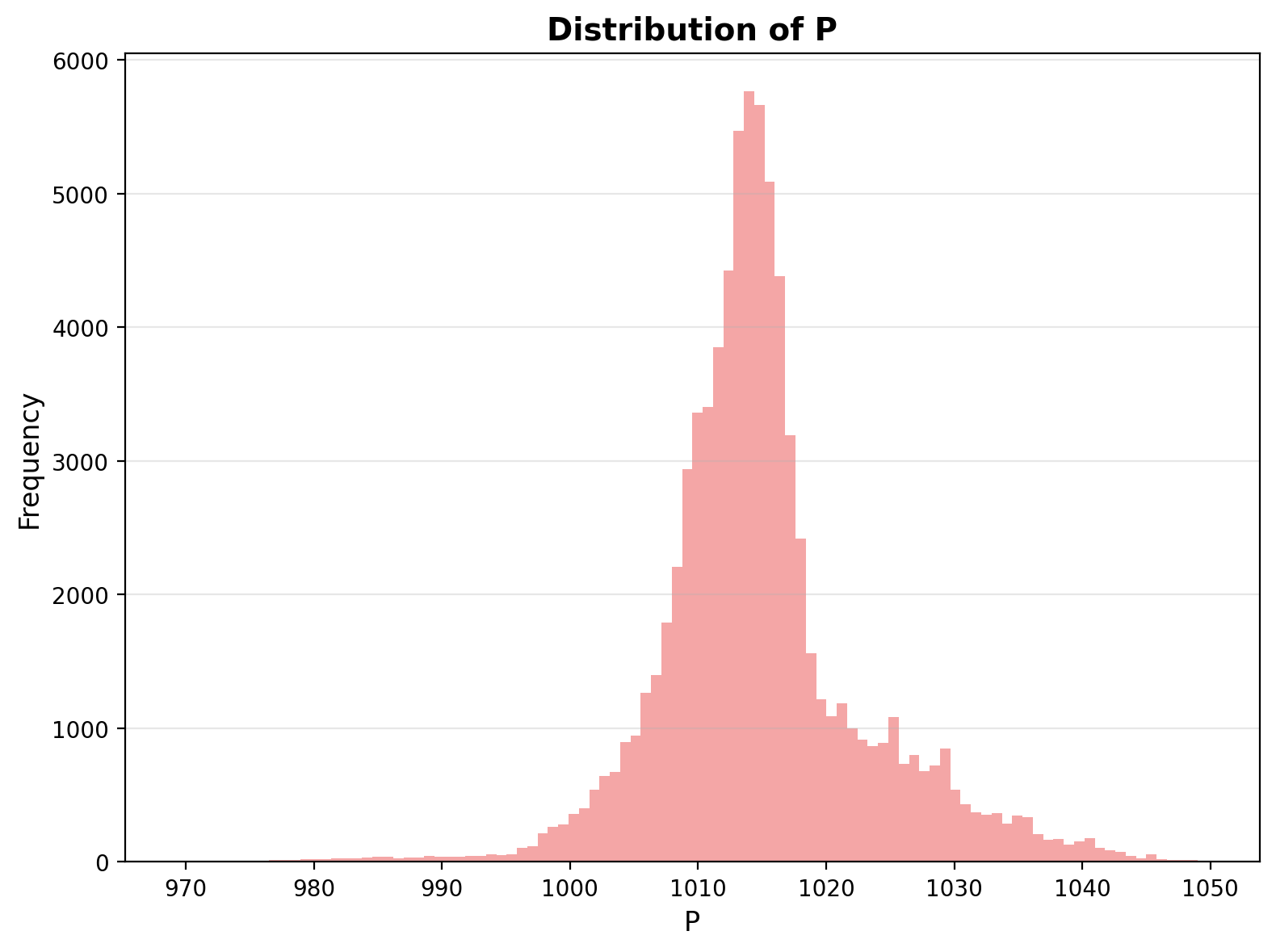}
        \caption{Histogram of values of \texttt{P} (pressure) in mb.}
    \end{subfigure}
    \hfill
    \begin{subfigure}[b]{0.35\textwidth}
        \includegraphics[width=\textwidth]{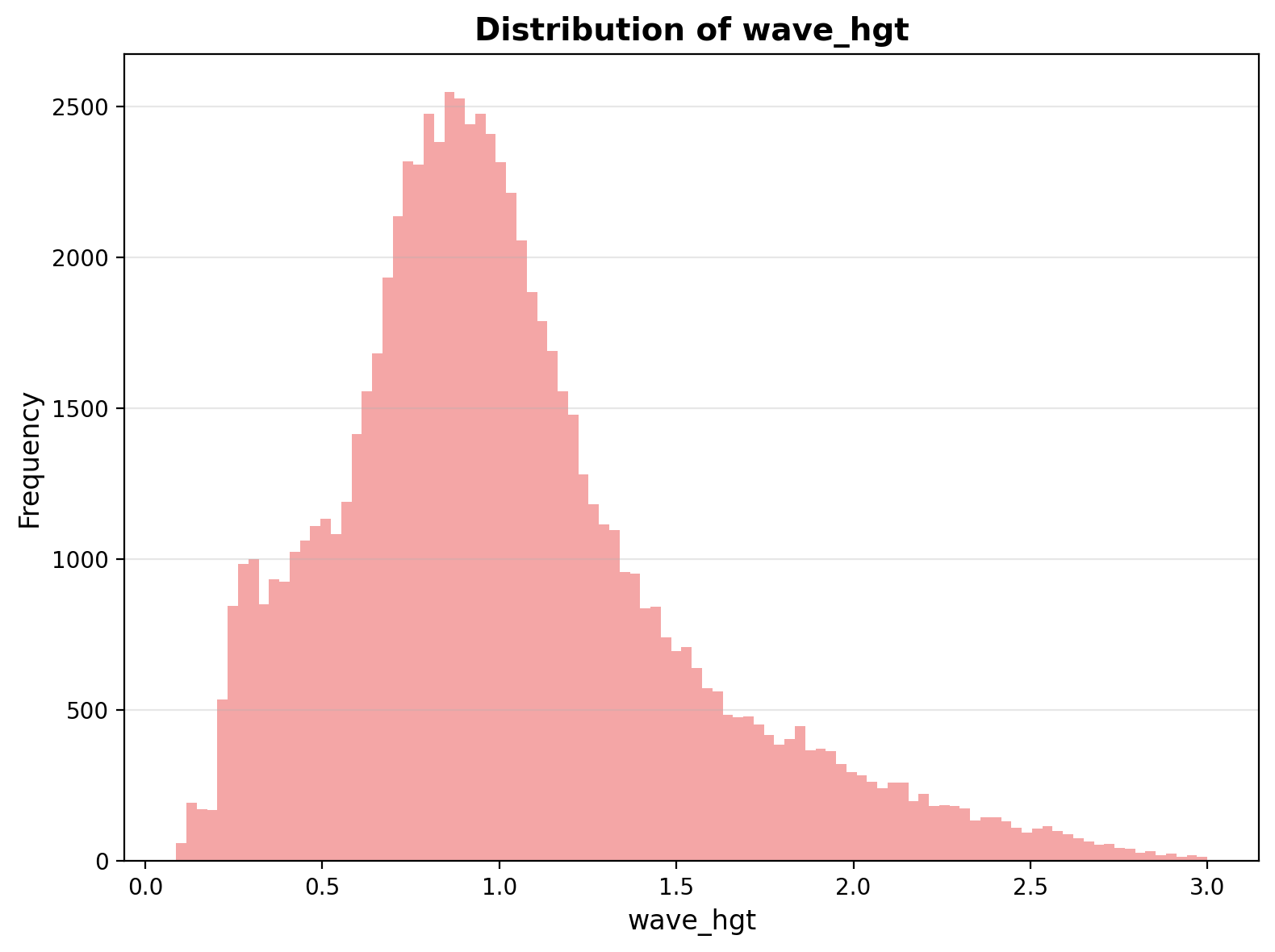}
        \caption{Histogram of values of \texttt{wave\_hgt} (wave height) in metres.}
    \end{subfigure}

    \caption{Histogram plots of some features.}
    \label{fig:histograms1}
\end{figure}

\begin{figure}[htbp]
    \centering

    \begin{subfigure}[b]{0.35\textwidth}
        \includegraphics[width=\textwidth]{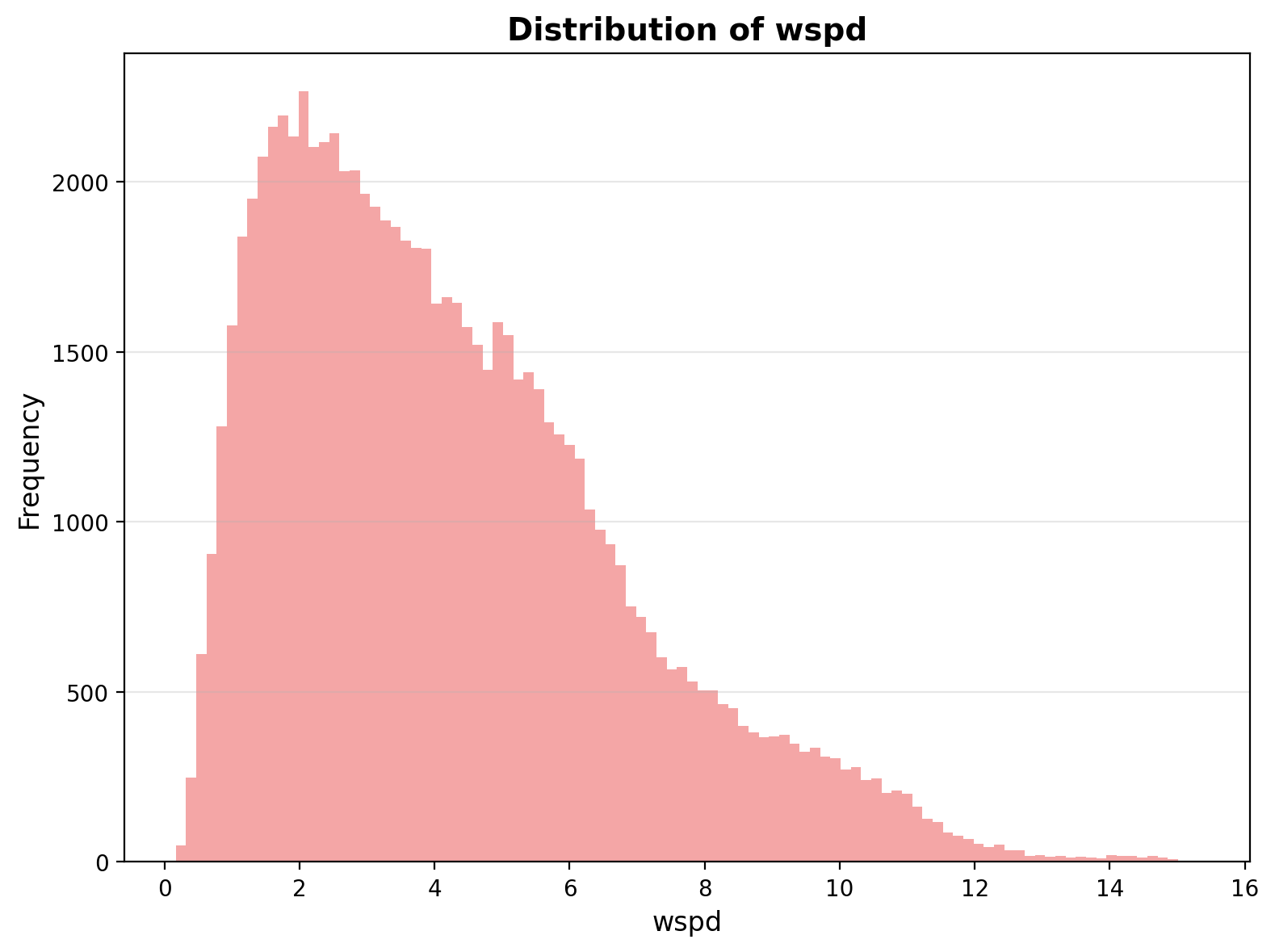}
        \caption{Histogram of values of \texttt{wspd} (wind speed) in m/s.}
    \end{subfigure}
    \hfill
    \begin{subfigure}[b]{0.35\textwidth}
        \includegraphics[width=\textwidth]{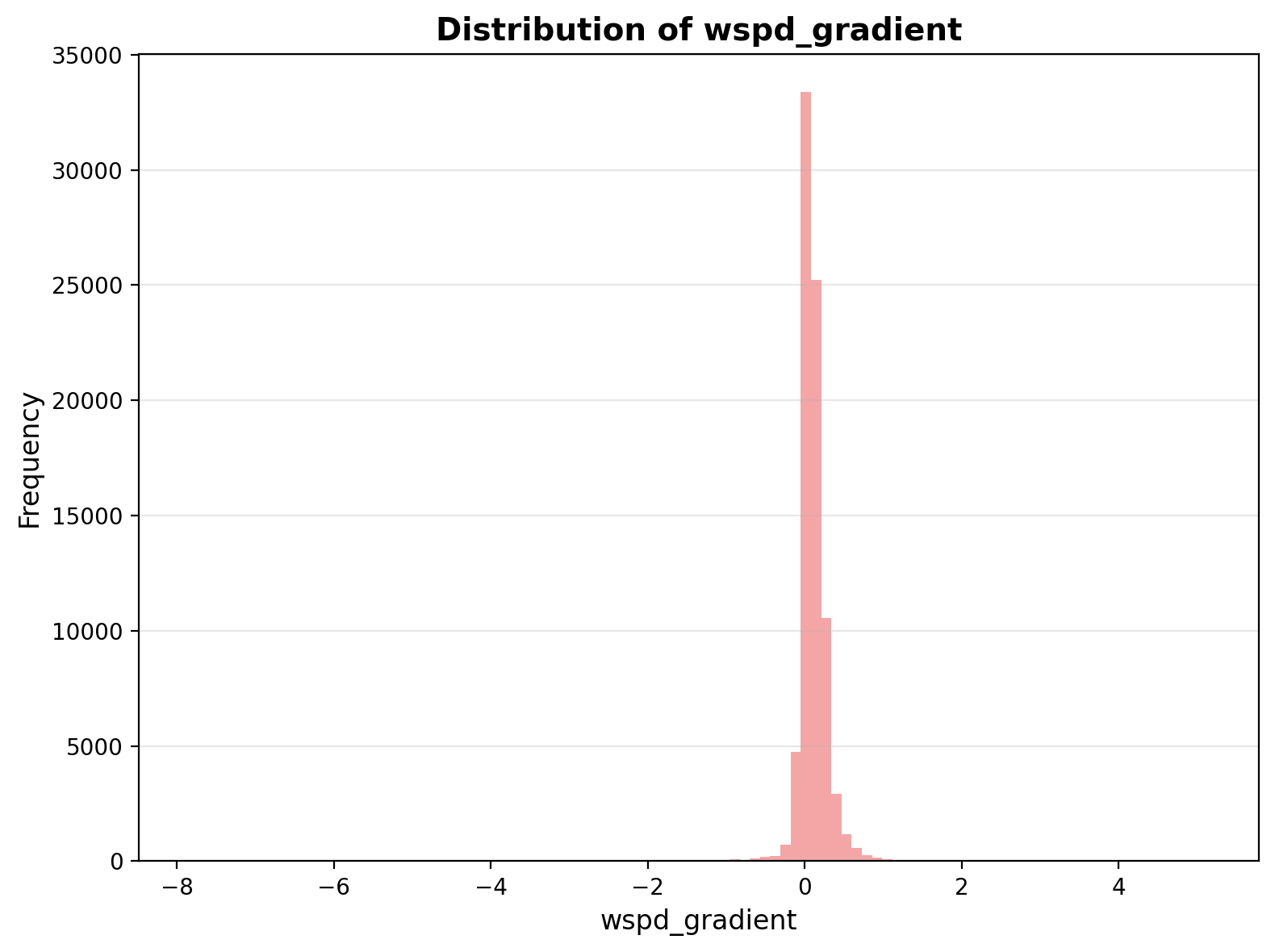}
        \caption{Histogram of values of \texttt{wspd\_gradient} (wind speed gradient) in m/s per metre.}
    \end{subfigure}

    \vspace{1em}

    \begin{subfigure}[b]{0.35\textwidth}
        \includegraphics[width=\textwidth]{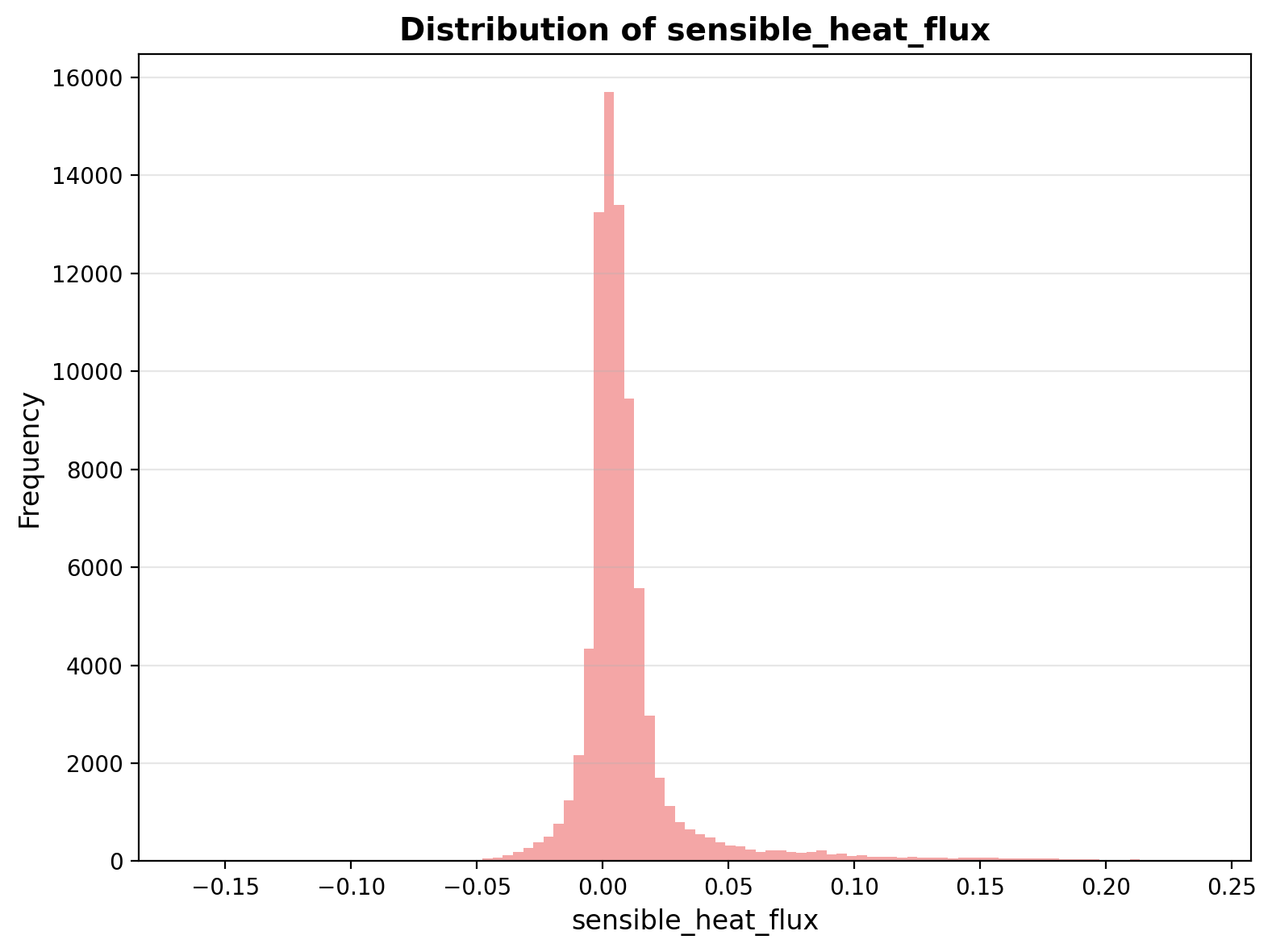}
        \caption{Histogram of values of \texttt{sensible\_heat\_flux} in K$\cdot$m/s.}
    \end{subfigure}
    \hfill
    \begin{subfigure}[b]{0.35\textwidth}
        \includegraphics[width=\textwidth]{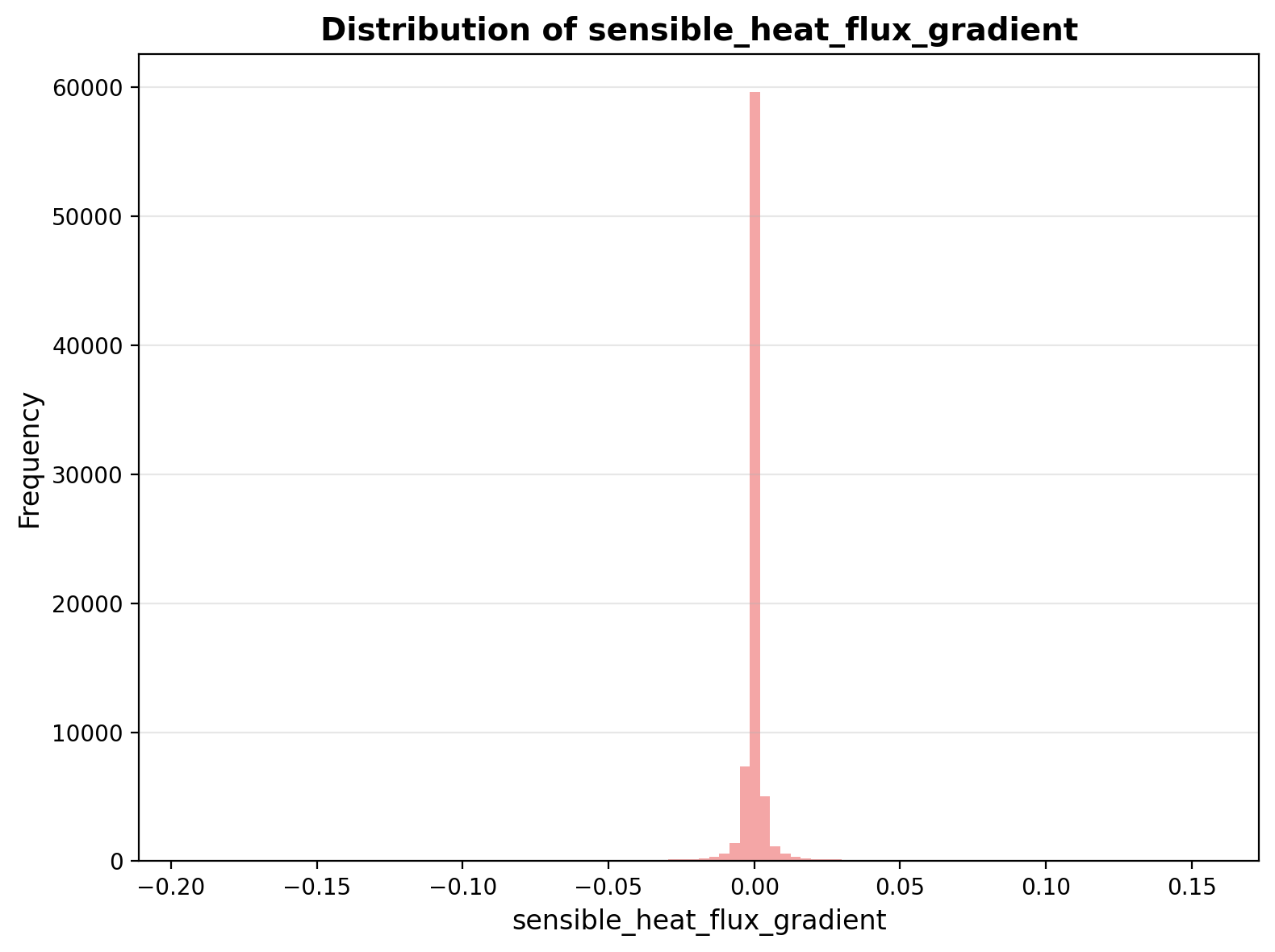}
        \caption{Histogram of values of \texttt{sensible\_heat\_flux\_gradient} in $K \cdot s^{-1}$.}
    \end{subfigure}

    \vspace{1em}

    \begin{subfigure}[b]{0.35\textwidth}
        \includegraphics[width=\textwidth]{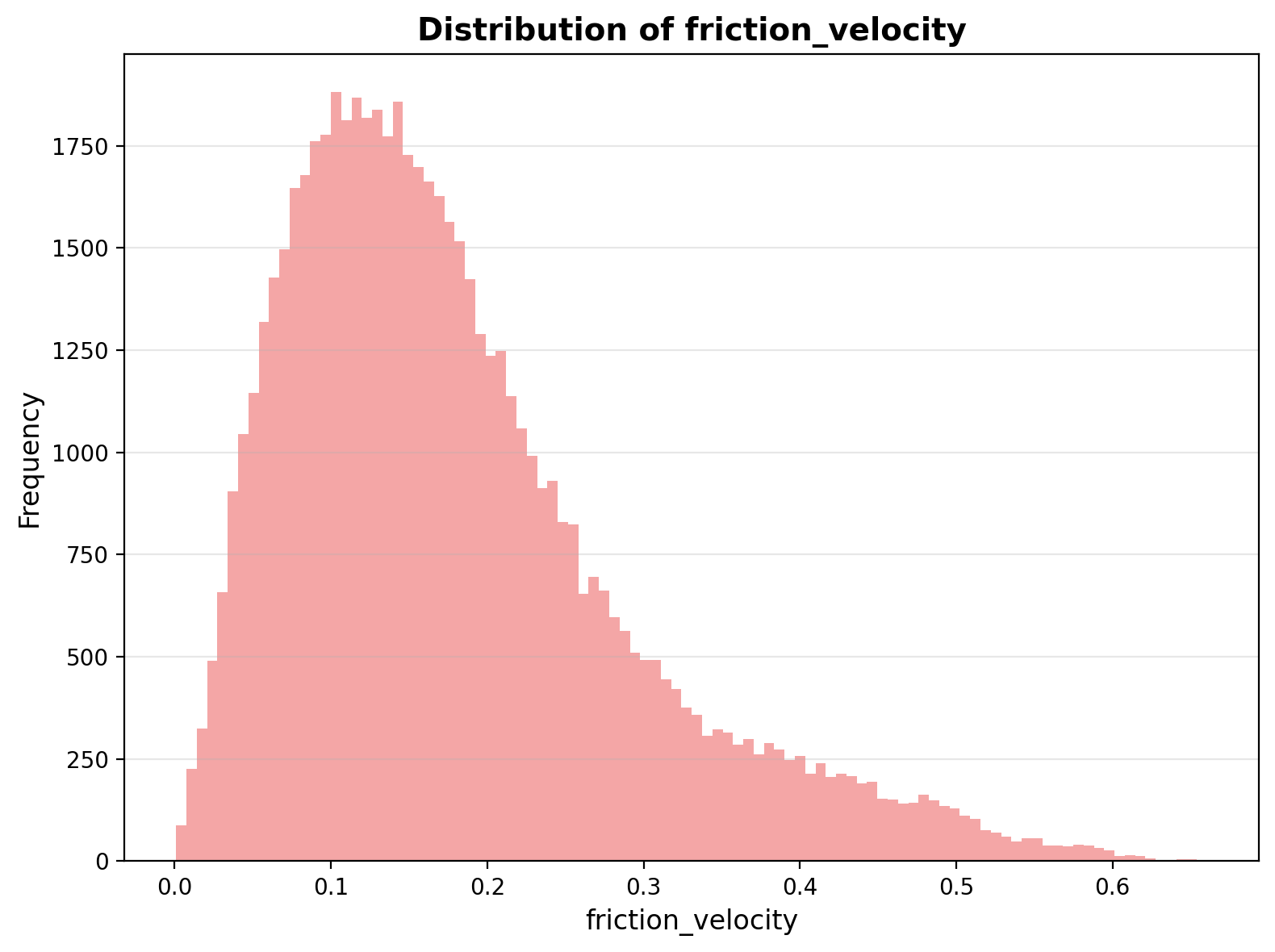}
        \caption{Histogram of values of \texttt{friction\_velocity} in m/s.}
    \end{subfigure}
    \hfill
    \begin{subfigure}[b]{0.35\textwidth}
        \includegraphics[width=\textwidth]{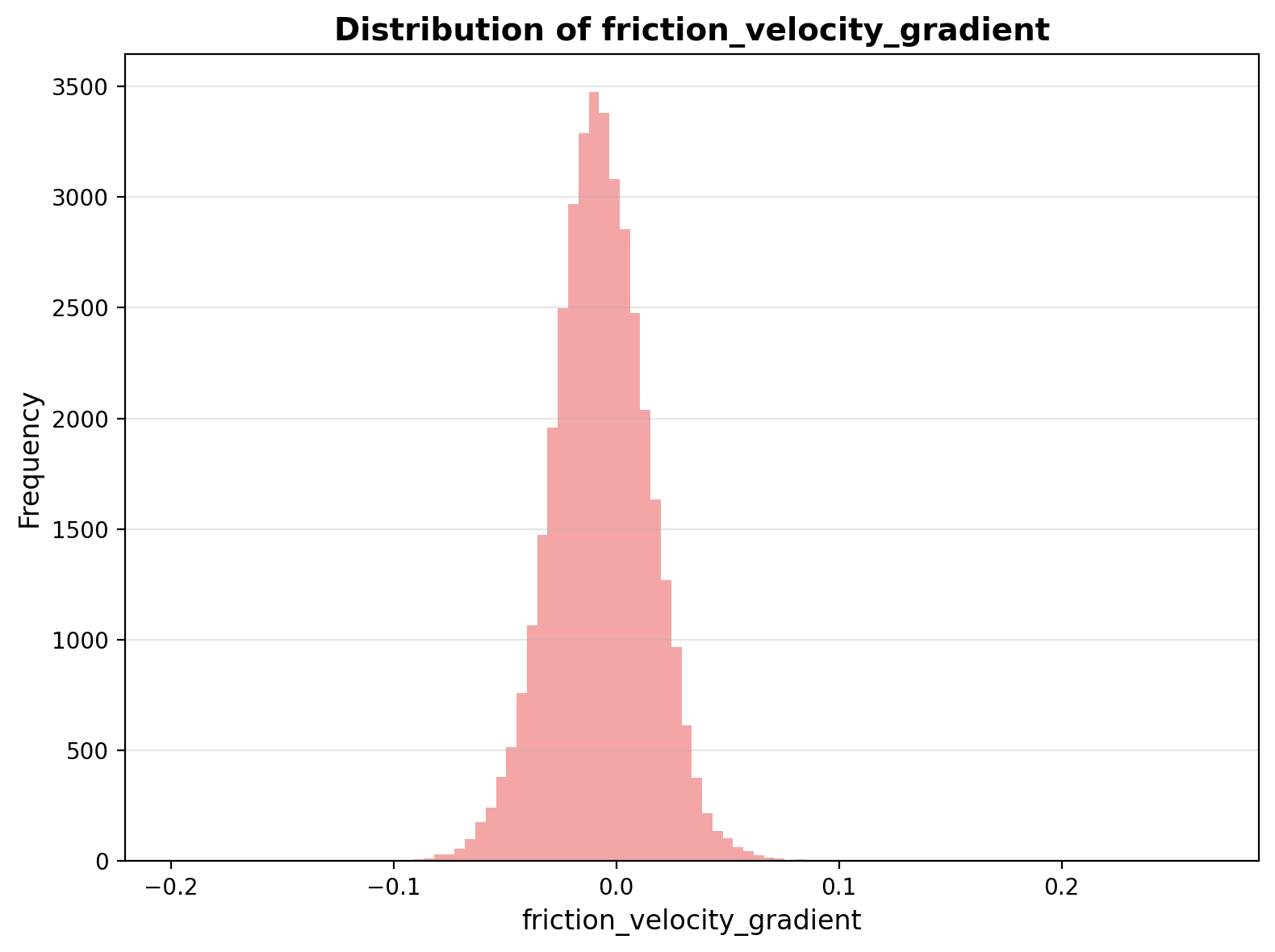}
        \caption{Histogram of values of \texttt{friction\_velocity\_gradient} in m/s per metre.}
    \end{subfigure}

    \caption{Histogram plots of some features.}
    \label{fig:histograms2}
\end{figure}

\begin{figure}[htbp]
    \centering

    \begin{subfigure}[b]{0.35\textwidth}
        \includegraphics[width=\textwidth]{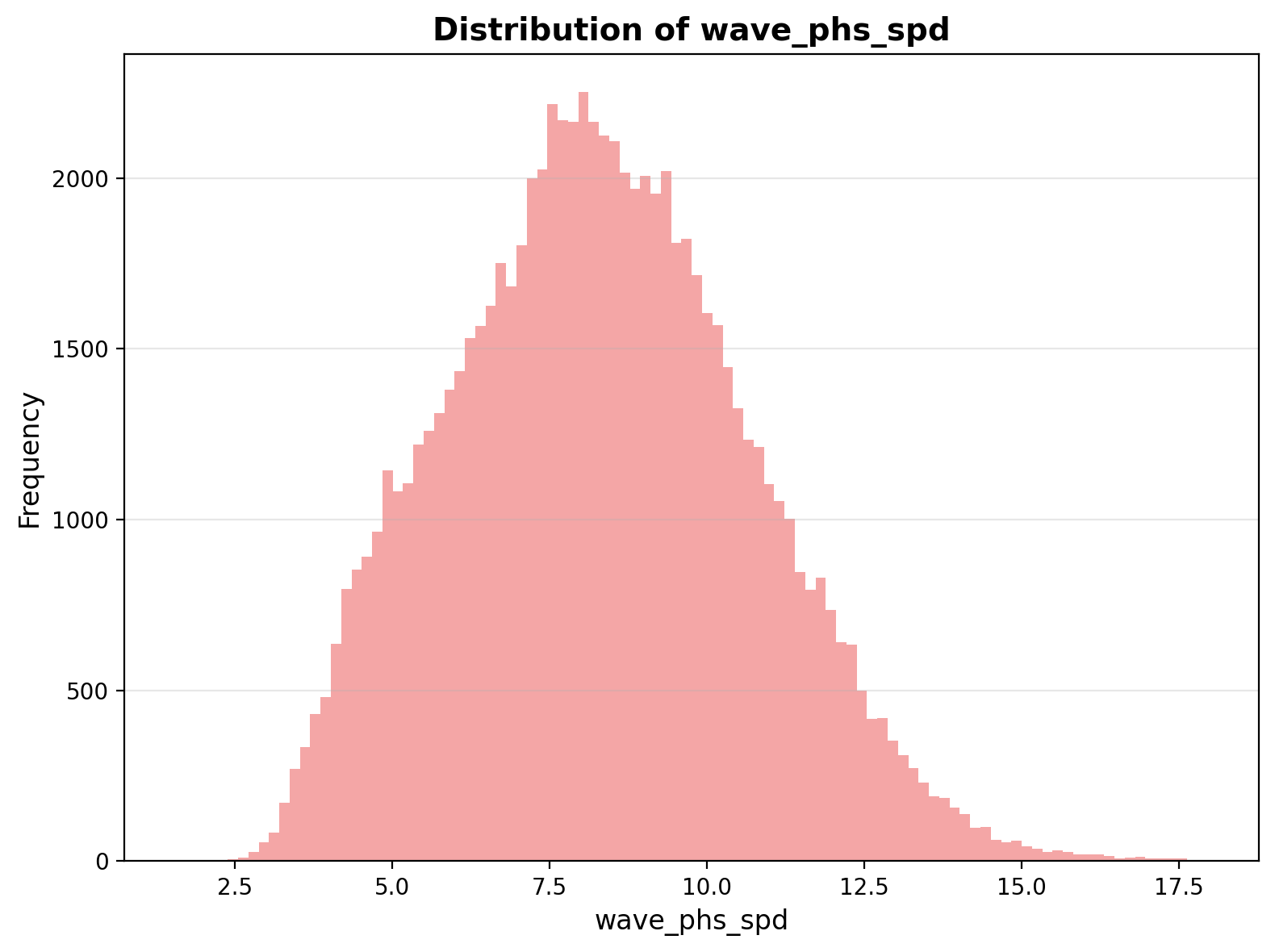}
        \caption{Histogram of values of \texttt{wave\_phs\_spd} (wave phase speed) in m/s.}
    \end{subfigure}
    \hfill
    \begin{subfigure}[b]{0.35\textwidth}
        \includegraphics[width=\textwidth]{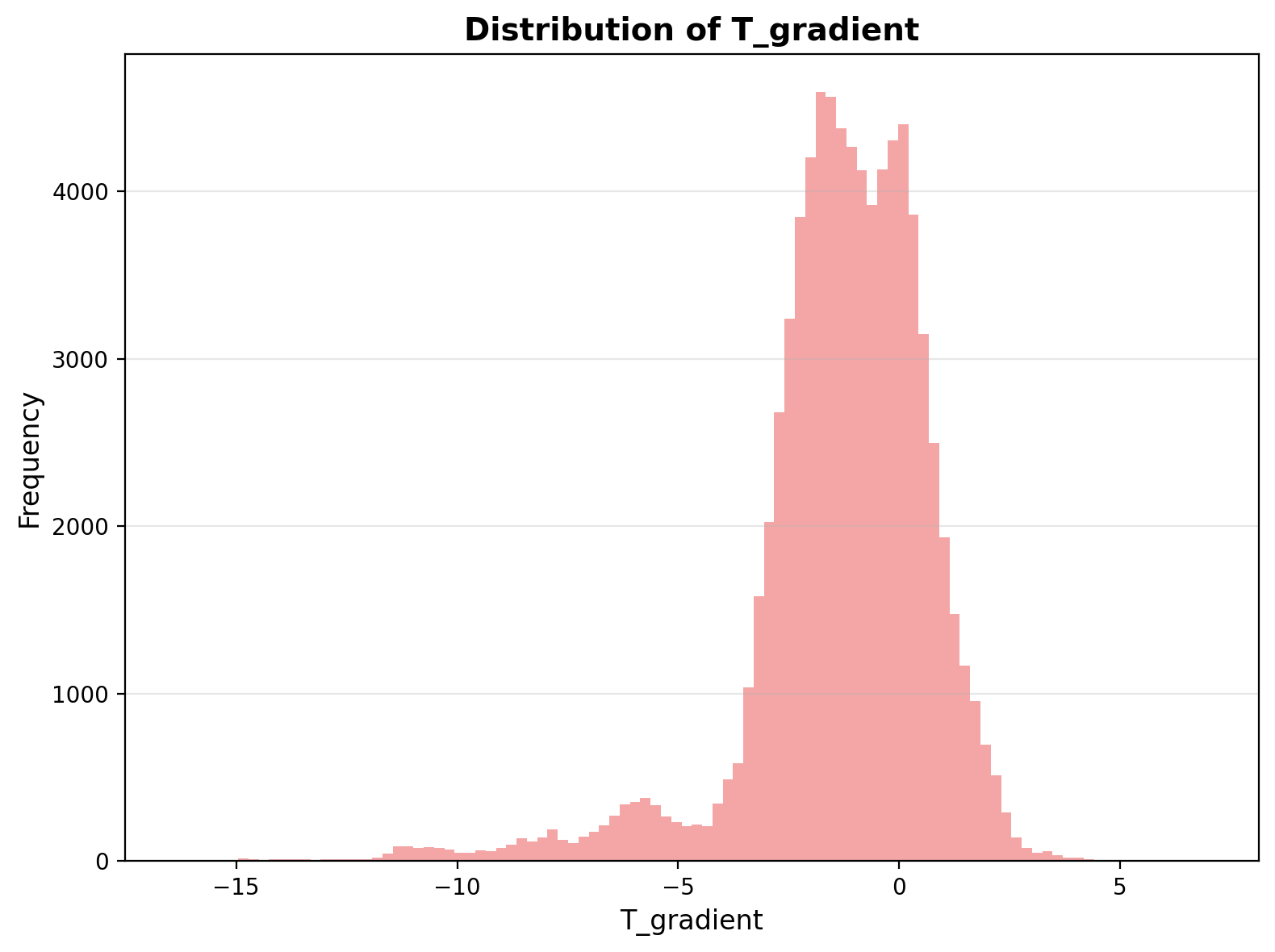}
        \caption{Histogram of values of \texttt{T\_gradient} (difference between air temperature and sea surface temperature) in Kelvin.}
    \end{subfigure}

    \vspace{1em}

    \begin{subfigure}[b]{0.35\textwidth}
        \includegraphics[width=\textwidth]{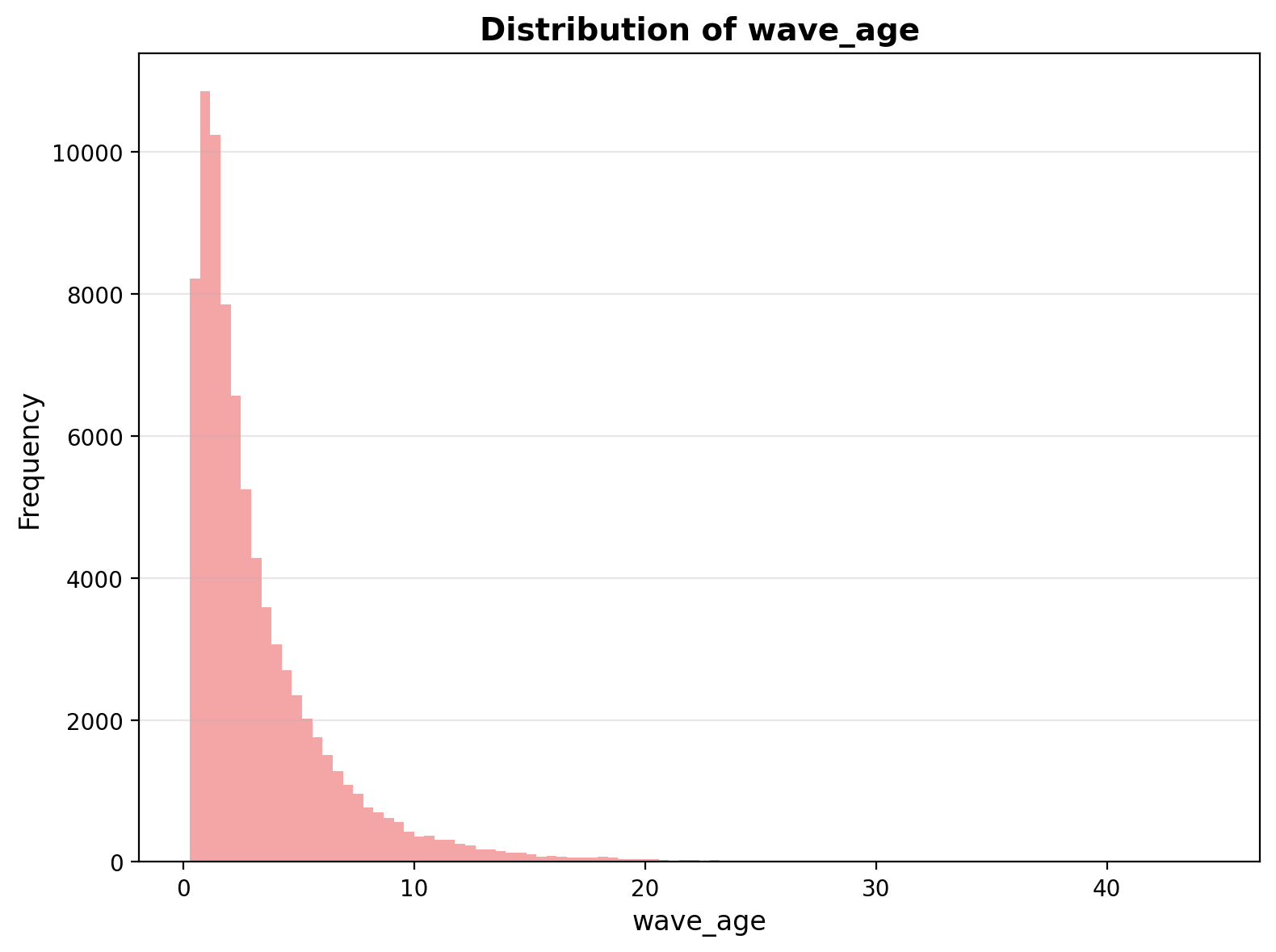}
        \caption{Histogram of values of \texttt{wave\_age}.}
    \end{subfigure}
    \hfill
    \begin{subfigure}[b]{0.35\textwidth}
        \includegraphics[width=\textwidth]{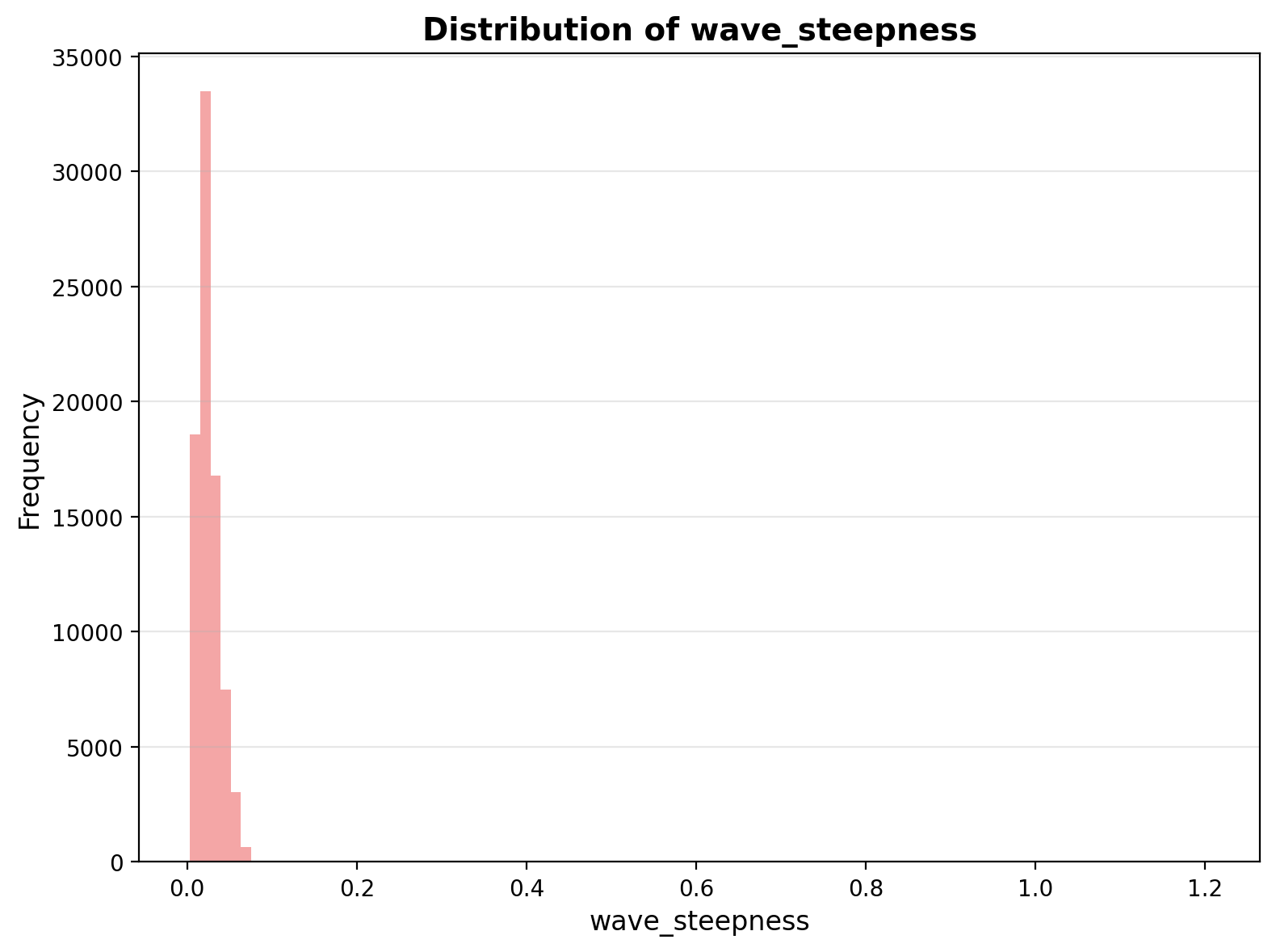}
        \caption{Histogram of values of \texttt{wave\_steepness}.}
    \end{subfigure}

    \vspace{1em}

    \begin{subfigure}[b]{0.35\textwidth}
        \includegraphics[width=\textwidth]{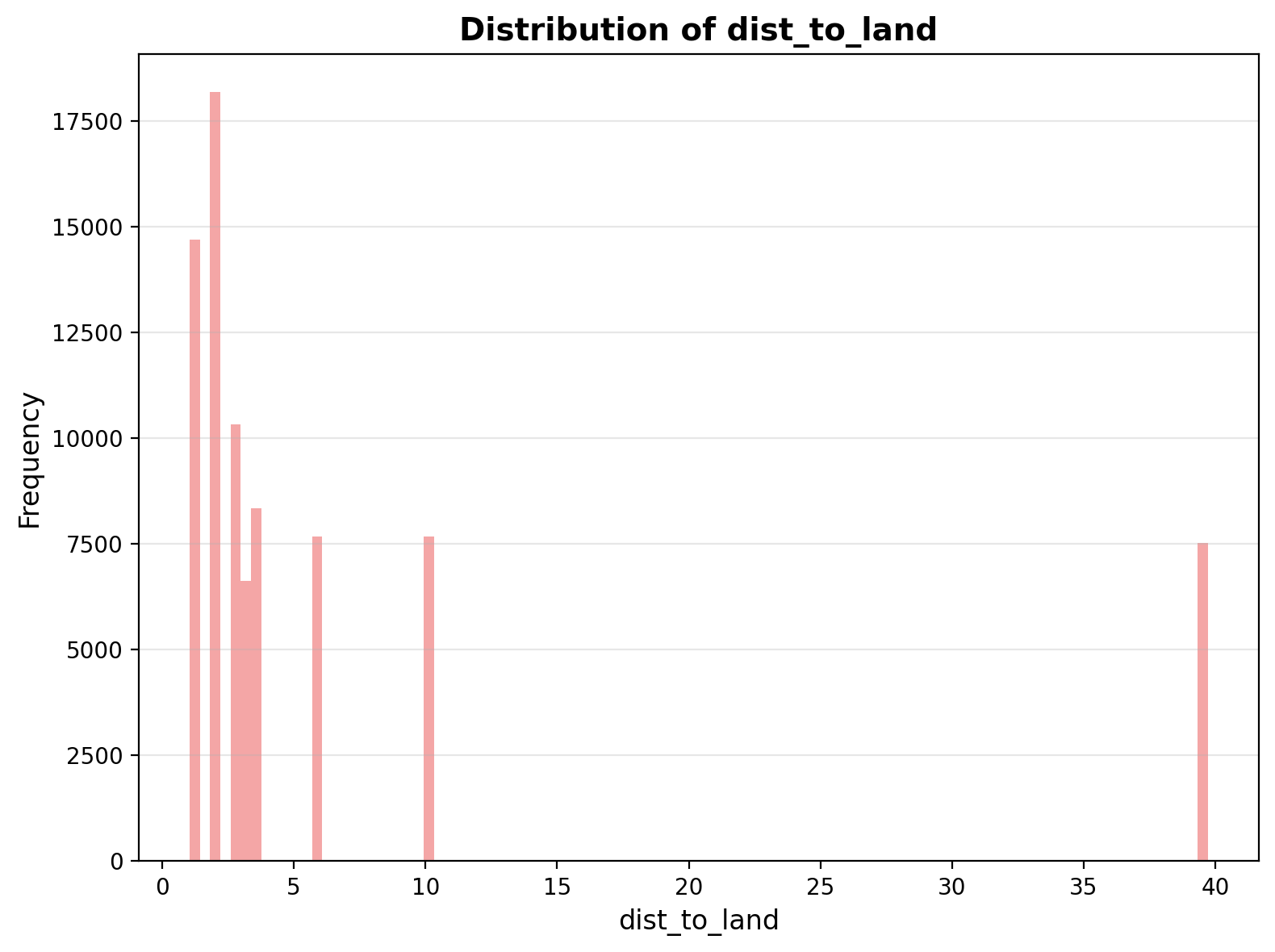}
        \caption{Histogram of values of \texttt{dist\_to\_land} (distance to land) in km.}
    \end{subfigure}
    \hfill
    \begin{subfigure}[b]{0.35\textwidth}
        \includegraphics[width=\textwidth]{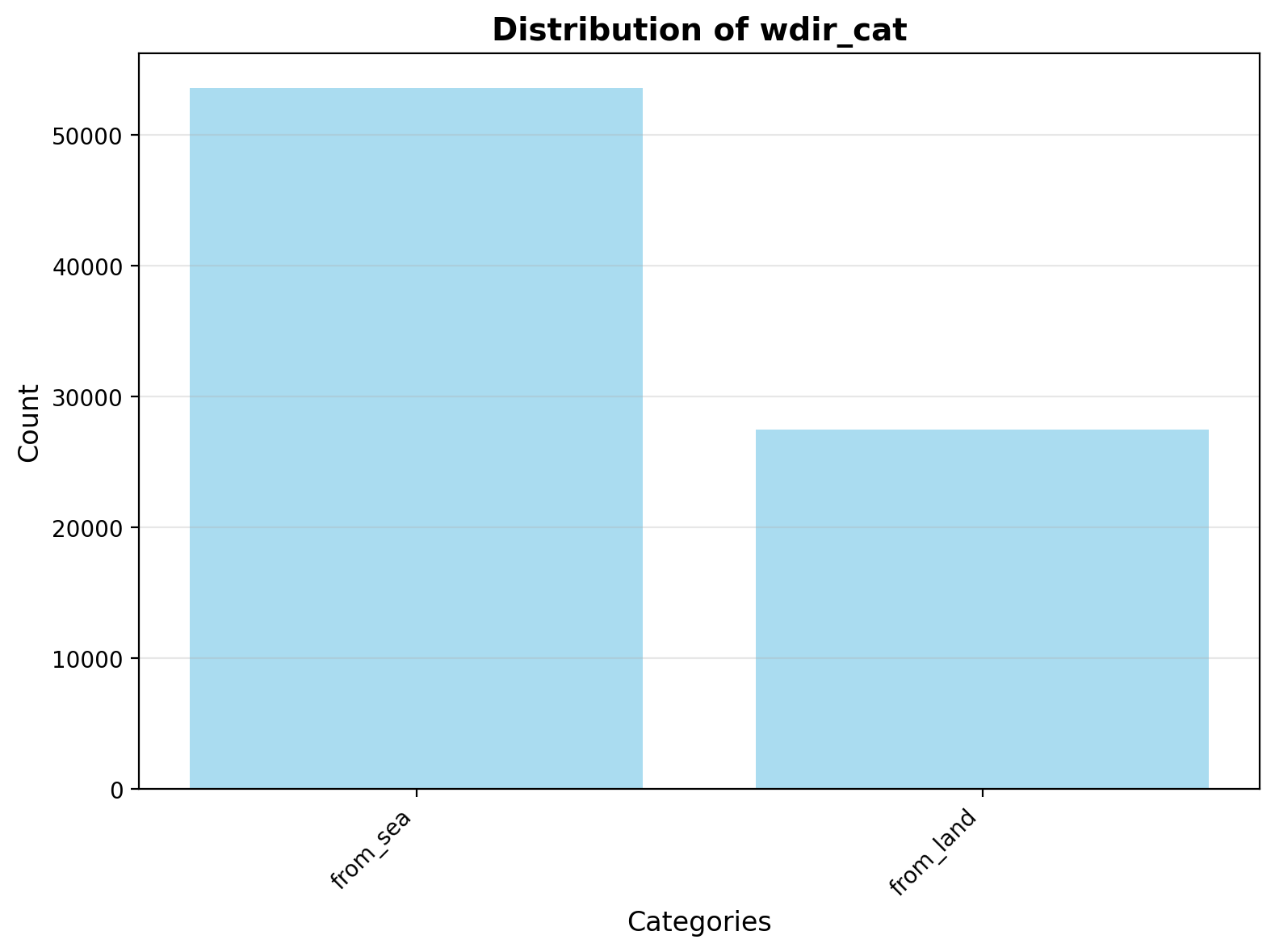}
        \caption{Bar plot of values of \texttt{wdir\_cat} (categorical transformation of wind direction, either from land or from sea).}
    \end{subfigure}

    \caption{Histogram plots of some features.}
    \label{fig:histogram3}
\end{figure}

\newpage
\section{Correlation and Mutual Information Matrices}

\begin{figure}[h]
    \centering
    \includegraphics[width=\linewidth]{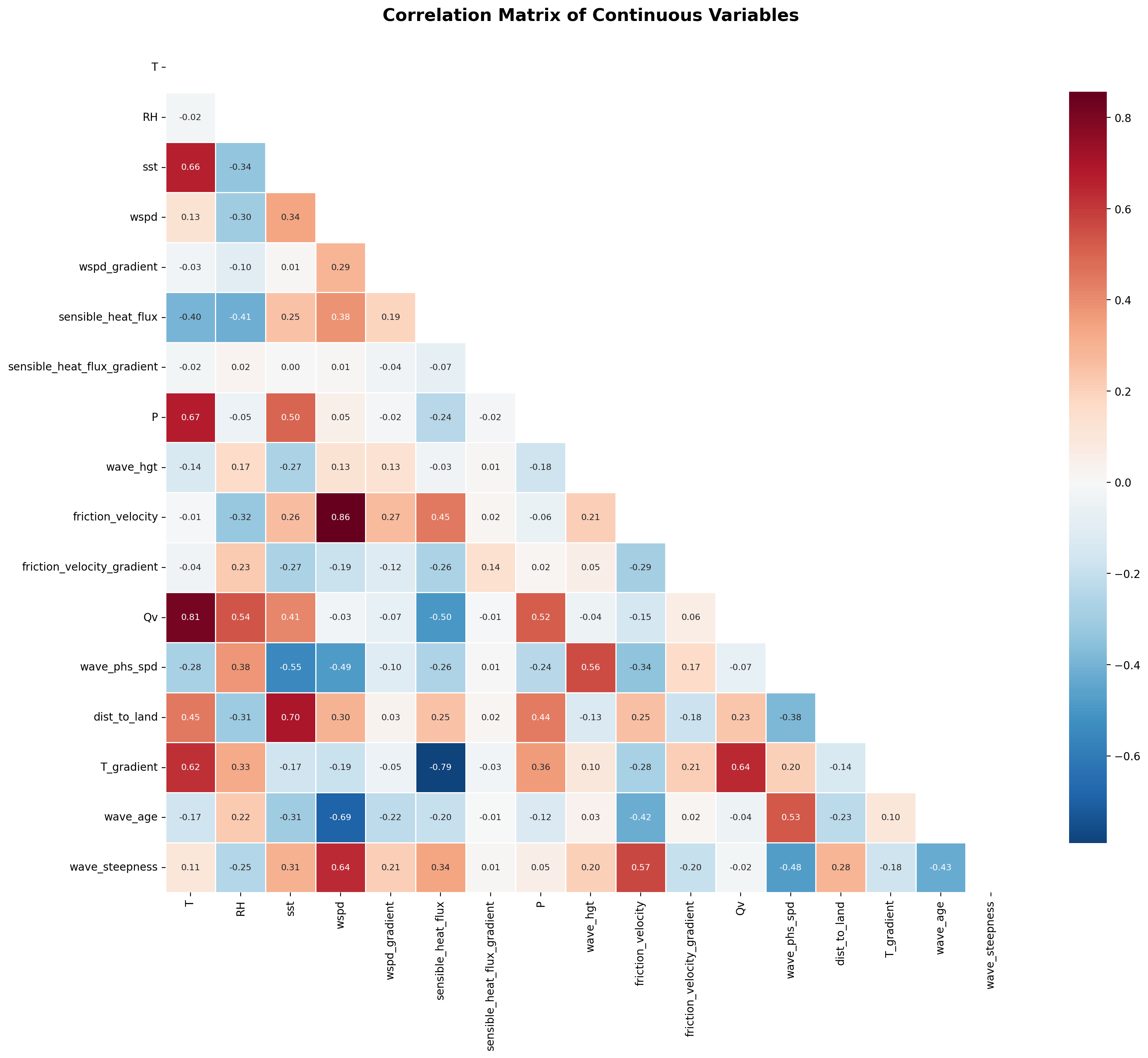}
    \caption{Correlation Matrix for Continuous Variables}
    \label{fig:corr-matrix}
\end{figure}

\begin{figure}[h]
    \centering
    \includegraphics[width=\linewidth]{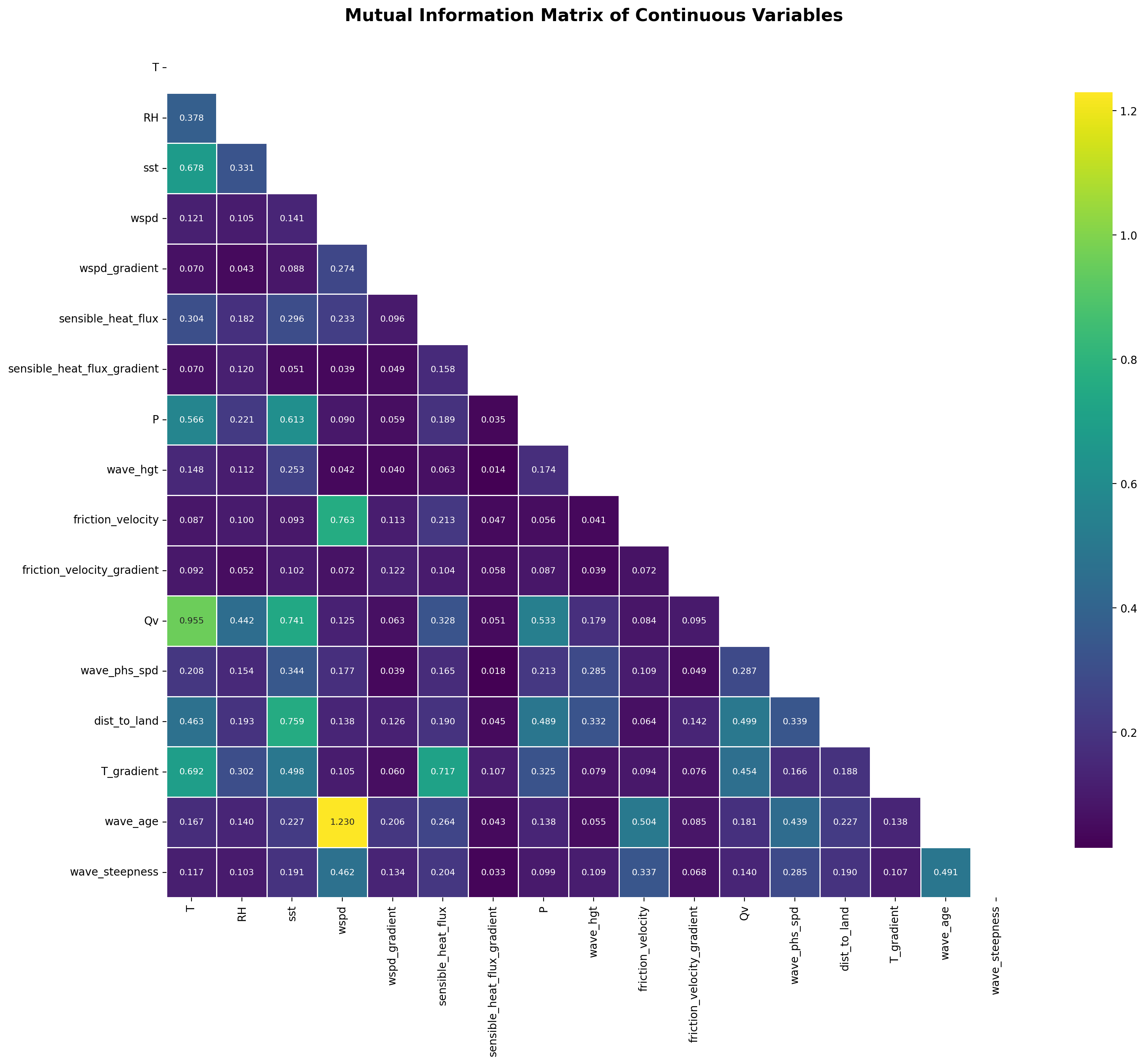}
    \caption{Mutual Information Matrix for Continuous Variables}
    \label{fig:mi-matrix}
\end{figure}

\end{document}